\documentclass[varenna]{cimento}

\usepackage{graphicx,epsfig}

\usepackage{dcolumn}     

\usepackage{amssymb}   

\def\beq{\begin{equation}}
\def\eeq{\end{equation}}
\newcommand{\expect}[1]{\langle {#1} \rangle}  
\newcommand{\case}{\frac}
\newcommand{\NN}{$N\!N~$}
\newcommand{\NNN}{$N\!N\!N~$}
\newcommand{\boldsigma}{\mbox{\boldmath$\sigma$}}

\title{Quantum Monte Carlo Calculations of Light Nuclei}

\author{Steven C. Pieper}
\institute{Physics Division \\ Argonne National Laboratory \\
Argonne, IL 60439, USA}

\shortauthor{Steven C. Pieper}

\begin{document}

\maketitle

\begin{abstract}
During the last 15 years, there has been much progress in
defining the nuclear Hamiltonian and
applying quantum Monte Carlo methods to the calculation
of light nuclei.
I describe both aspects of this work and some recent results.
\end{abstract}

\section{Introduction}

The goal of {\it ab-initio} light-nuclei calculations is
to understand nuclei as collections of nucleons
interacting with realistic (bare) potentials through
reliable solutions of the many-nucleon Schr\"{o}dinger equation.
Such calculations can study
binding energies, excitation spectra, relative stability,
densities, transition amplitudes, cluster-cluster overlaps,
low-energy astrophysical reactions, and other aspects of nuclei.
Such calculations are also essential to claims of sub-nucleonic
effects, such as medium modifications of the nuclear force or nucleon
form factors; if a reliable pure nucleonic degrees of freedom calculation
can reproduce experiment then there is no basis for
claims of seeing sub-nucleonic degrees
of freedom in that experiment (beyond the obvious fact that the
free-space nucleon interactions are a result of sub-nucleonic degrees
of freedom).

There are two problems in microscopic few- and many-nucleon
calculations: 1) determining the Hamiltonian, and 2) given $H$,
accurately solving the Schr\"{o}dinger equation for $A$ nucleons; I will
discuss both of these in this contribution.  The two-nucleon (\NN$\!$) force is
determined by fitting the large body of $N\!N$ scattering data.  Several
modern $N\!N$ potentials are in common use.  The Argonne $v_{18}$ is a local
potential written in operator format; this potential is used in the
calculations described here, and is presented in some detail below.
Other modern potentials are generally non-local; some of them are
discussed in other contributions to this school.  

It has long been known
that calculations with just realistic $N\!N$ potentials fail to
reproduce the binding energies of nuclei; three-nucleon (\NNN$\!$) potentials are
also required.  These arise naturally from an underlying meson-exchange
picture of the nuclear forces or from chiral effective field theories.
Unfortunately, much \NNN scattering data is well reproduced by
calculations using just $N\!N$ forces, so the $N\!N\!N$ force must
determined from properties of light nuclei.  In this contribution
the recent Illinois models with $2\pi$ and $3\pi$
rings are used.

Our understanding of nuclear forces has evolved over the last 70 years:
\begin{itemize}
\renewcommand{\itemsep}{-4pt}
\item Meson-exchange theory of Yukawa (1935)
\item Fujita-Miyazawa three-nucleon potential (1955)
\item First phase-shift analysis of $N\!N$ scattering data (1957) 
\item Gammel-Thaler, Hamda-Johnston and Reid phenomenological potentials (1957--1968) 
\item Bonn, Nijmegen and Paris field-theoretic models (1970s)
\item Tuscon-Melbourne and Urbana $N\!N\!N$ potential models (late 70's--early 80's)
\item Nijmegen partial wave analysis (PWA93) with $\chi^2$/dof$\sim1$ (1993)
\item Nijm~I, Nijm~II, Reid93, Argonne v$_{18}$ and CD-Bonn (1990s)
\item Effective field theory at N$^3$LO (2004)
\end{itemize}
References for a number of these developments are given in
the following sections.

Accurate solutions of the many-nucleon Schr\"{o}dinger equation
have also evolved over many decades:
\begin{itemize}
\renewcommand{\itemsep}{-4pt}
\item $^2$H  by  numerical integration (1952) -- a
pair of coupled second-order differential equations in 1 variable.
At the time this took
``between 5 and 20 minutes for the calculation and the printout 
another 5 minutes''~\cite{BK53}!
\item $^3$H  by  Faddeev (1975--1985)
\item $^4$He by  Green's function Monte Carlo (GFMC) (1988)
\item $A=6$  by  GFMC and No-core shell model (NCSM) (1994-95)
\item $A=7$  by  GFMC and NCSM (1997-98)
\item $A=8$  by  GFMC and NCSM (2000)
\item $^4$He benchmark by 7 methods to 0.1\% (2001)
\item $A=9,10$  by  GFMC and NCSM (2002)
\item $^{12}$C by  GFMC and NCSM (2004--)
\item $^{16}$O by  Coupled Cluster (CC) (2005--)
\end{itemize}
References for the $A$=3,4 calculations may be found in Ref.~\cite{CS98};
the GFMC calculations are the subject of this paper; the NCSM are
discussed in Petr Navr\'atil's contribution to this Course; and
CC results may be found in Ref.~\cite{coup-clus-07}.

This contribution is limited to Variational Monte Carlo (VMC)
and GFMC calculations of light nuclei.  Section~\ref{sec:ham}
describes the Hamiltonians used and sections~\ref{sec:qmc} 
through~\ref{sec:gfmc} describe the computation methods.
Section~\ref{sec:energy} gives a number of results for energies
of nuclear states; Sec.~\ref{sec:scat} describes GFMC
calculations of scattering states; and Sec.~\ref{sec:dens}
gives some results for densities.  Finally some conclusions
and prospects for the future are presented in Sec.~\ref{sec:conclusions}.

\section{Hamiltonians}
\label{sec:ham}

The nuclear Hamiltonian used here has the form
\begin{equation}
H = { \sum_{i} K_i } + { { \sum_{i<j}} v_{ij} } 
+ { \sum_{i<j<k} V_{ijk} } \ .
\end{equation}
Here $K_i$ is the non-relativistic kinetic energy, including
$m_n-m_p$ effects, $v_{ij}$ is the \NN potential and $V_{ijk}$ is
the \NNN potential.

\subsection{Argonne $v_{ij}$}

The $N\!N$ potential ($v_{ij}$) is  Argonne v18~\cite{WSS95} (AV18)
which has the form
\begin{equation}
   { v_{ij}} = { v^{\gamma}_{ij}} + { v^{\pi}_{ij}} + { v^{R}_{ij}} + { v^{CIB}_{ij}} .
\end{equation}
The $v^{\gamma}_{ij}$ is a very complete representation of
the $pp$, $pn$ and $nn$ electromagnetic terms, including
first- and second-order Coulomb, magnetic, vacuum polarization, etc., components
with form factors.
(Ref.~\cite{W97} provides a heuristic introduction to AV18.)

The $v^{\pi}_{ij}$ is the isoscalar one-pion exchange potential represented 
as a local operator:
\begin{eqnarray}
v^{\pi}_{ij}  &=& \frac{1}{3} \  \frac{f^2_{\pi NN}}{4 \pi} \  m_{\pi} \ 
X_{ij} \ {\bf\tau}_{i}\cdot{\bf\tau}_{j} \ ,
\label{eq:vpi} \\
X_{ij} &=& T(m_{\pi}r_{ij}) \  S_{ij} + Y(m_{\pi}r_{ij}) \  
\boldsigma_i \cdot \boldsigma_j \ , 
\label{eq:xij}    \\
Y(x) &=& \frac{e^{-x}}{x} \  \xi(r) \ , 
\label{eq:yij}   \\
T(x) &=& \left( \frac{3}{x^2} + \frac{3}{x} + 1 \right) Y(x) \  \xi(r) \ ,
\label{eq:tij}  \\
\xi(r)  &=& (1- e^{-c_\pi r^2}) \ .
\end{eqnarray}
where ${\bf\tau}_{i}$, $\boldsigma_i$ and $S_{ij}$ are isospin, spin and
tensor operators, respectively.
In light nuclei, $\langle v^{\pi}_{ij} \rangle$ contributes $\sim$85\% of $\langle v_{ij} \rangle$.

The remaining isospin-conserving terms are
\begin{eqnarray}
v^{R}_{ij} &=& \sum_{p=1,14} v_{p}(r_{ij}) O^{p}_{ij} \ , \\
O^{p=1,14}_{ij} &=& [1, {\bf\sigma}_{i}\cdot{\bf\sigma}_{j}, S_{ij},
{\bf L\cdot S},{\bf L}^{2},{\bf L}^{2}{\bf\sigma}_{i}\cdot{\bf\sigma}_{j},
({\bf L\cdot S})^{2}]\otimes[1,{\bf\tau}_{i}\cdot{\bf\tau}_{j}] \ , 
\end{eqnarray}
where $v_{p}(r)$ has short-, intermediate-, and long-range components.
The long-range components are just the $Y(r)$ and $T(r)$ of the 
one-pion potential and are present only for those operators that have
contributions from one-pion exchange.  The intermediate-range components
are proportional to $T^2(r)$ and the short-range component is of
the Woods-Saxon form.  

Finally, $v^{CIB}_{ij}$ is the strong charge independence breaking 
part of the potential and consists of four operators:
\begin{equation}
   O^{p=15,18}_{ij} = [1, ({\bf\sigma}_{i}\cdot{\bf\sigma}_{j}), 
S_{ij}]\otimes T_{ij} , (\tau_{zi}+\tau_{zj}) \ .
\end{equation}
The long-range part of $O^{p=15,17}$ comes from one-pion exchange
by inserting $m_{\pi^{+-}}$ or $m_{\pi^0}$ in Eqn.~(\ref{eq:vpi}
and \ref{eq:xij}) and using ${f^2_{\pi NN}} \propto m_{\pi}$.

The parameters in the short- and intermediate-range components were
determined by making a direct fit to the 1993 Nijmegen data
base~\cite{BCKKRSS90,SKRS93} containing 1787 $pp$ and 2514 $np$ data in
the range $0-350$ MeV, the $nn$ scattering length, and deuteron binding
energy.  The fit of approximately 40 parameters results in a
$\chi^2/$d.o.f. of 1.09, which is typical of 1990's $N\!N$ potentials.

\subsection{Illinois $V_{ijk}$}

The three-nucleon potential used for most of the examples
presented here is the Illinois-2~\cite{PPWC01}.  It
consists of two- and three-pion terms and a simple phenomenological
repulsive term:
\begin{equation}
V_{ijk} = { V^{2\pi}_{ijk}} + { V^{3\pi}_{ijk}} + { V^{R}_{ijk}} \ .
\end{equation}

\begin{figure}[bt] 
\includegraphics[height=.75in]{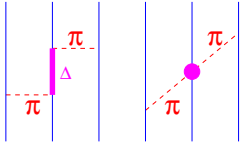}
\caption{Two-pion exchange terms in the Illinois \NNN potentials.}
\label{fig:2pi}
\end{figure}

The two-pion term, illustrated in Fig.~\ref{fig:2pi}, contains
$P$- and $S$-wave $\pi N$-scattering terms:
\begin{equation}
V^{2\pi}_{ijk} = V^{2\pi,P}_{ijk} + V^{2\pi,S}_{ijk}    \ .
\end{equation}
The $P$-wave term (left panel of Fig.~\ref{fig:2pi}) is the well-known 
Fujita-Miyazawa~\cite{FM57a,FM57b} term
which is present in all realistic \NNN potentials.  It has the form
\begin{equation}
   V^{2\pi,P}_{ijk} = A_{2\pi,P}  \sum_{cyclic} 
           \{X_{ij},X_{jk}\} 
           \{\tau_{i}\cdot\tau_{j},\tau_{j}\cdot\tau_{k}\}
         + \frac{1}{4} [X_{ij},X_{jk}]
           [\tau_{i}\cdot\tau_{j},\tau_{j}\cdot\tau_{k}] \ ,
\end{equation}
where $X_{ij}$ is defined in Eq.~(\ref{eq:xij}).  This is the
longest-ranged nuclear \NNN potential and is attractive in
all nuclei and nuclear matter.  However it is very small
or even slightly repulsive in pure neutron systems.

The second panel of Fig.~\ref{fig:2pi} represents the $S$-wave
part of $V^{2\pi}_{ijk}$.  This term was introduced in the
Tuscon-Melbourne \NNN potential~\cite{TM79} and is required by
chiral perturbation theory.  However, in practice it is
only 3\%--4\% of $V^{2\pi,P}_{ijk}$ in light nuclei.

\begin{figure}[bt] 
\includegraphics[height=.75in]{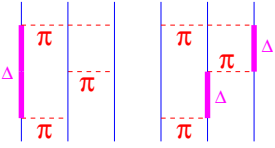}
\caption{Three-pion ring terms in the Illinois \NNN potentials.}
\label{fig:3pi}
\end{figure}

The three-pion term (Fig.~\ref{fig:3pi}) was introduced in the
Illinois potentials.  It consists of the subset of three-pion
rings that contain only one $\Delta$ mass in the energy denominators.
Even so it has a quite complicated form which is given in Ref.~\cite{PPWC01}.
An important aspect of this structure is that there is a significant
attractive term which acts only in $T$=3/2 triples.  In most light
nuclei 
$\langle V^{3\pi}_{ijk} \rangle \lesssim  0.1 \langle V^{2\pi}_{ijk} \rangle$

The final term in the \NNN potential, $V^{R}_{ijk}$, represents all other
diagrams including relativistic effects.  It is strictly phenomenological and
purely central and repulsive:
\begin{equation}
V^{R}_{ijk} = A_R \sum_{cyclic} T^2(m_{\pi}r_{ij}) T^2(m_{\pi}r_{jk}) \ .
\end{equation}
This repulsive term is principally needed to make nuclear
matter saturate at the proper density instead of a too-high
density and to obtain a hard enough equation of state for neutron matter.

The coupling constants $A_{2\pi,P}$, $A_{3\pi}$, and $A_R$ were
adjusted to fit 17 nuclear levels for $A \leq 8$.  The $V^{2\pi,S}_{ijk}$
is too weak to be determined by fitting and its coupling was left
at the value predicted by chiral perturbation theory.

In light nuclei we find 
\begin{equation}
\expect{V_{ijk}}  \sim  (0.02 ~\mbox{to}~ 0.09)\expect{v_{ij}} 
                      \sim  (0.15 ~\mbox{to}~ 0.6)\expect{H} 
\end{equation}
where the large fraction of $\expect{H}$ is due to a large cancellation 
of $K$ and $v_{ij}$.  From this we expect
\begin{equation}
\expect{V_{4N}} \sim 0.06 \expect{V_{ijk}} \
                \sim  (0.02 ~\mbox{to}~ 0.04)\expect{H} 
                \sim  (0.5 ~\mbox{to}~ 2.)~ \mbox{MeV} \ .
\end{equation}
This is comparable to the accuracy of our calculations.  Even if
more accurate calculations could be made, it  would probably not 
be possible to disentangle four-nucleon potential effects from 
uncertainties in the fitted parameters of $V_{ijk}$.

\subsection{What Makes Nuclear Structure?}
\label{sec:avx}

We have defined a very complicated nuclear Hamiltonian and it is
reasonable to ask if it is all necessary to reproduce the structure
of light nuclei.  A study~\cite{WP02} was made of this in which
features of the nuclear Hamiltonian were systematically removed
and the effects on nuclear level energies investigated.  For
each simplification of the two-nucleon part of $H$, the
remaining terms were readjusted to continue reproducing
as many low partial-wave phase shifts, and the deuteron, as possible.

\begin{figure}[tb] 
\centering
\includegraphics[height=5.30in,angle=270]{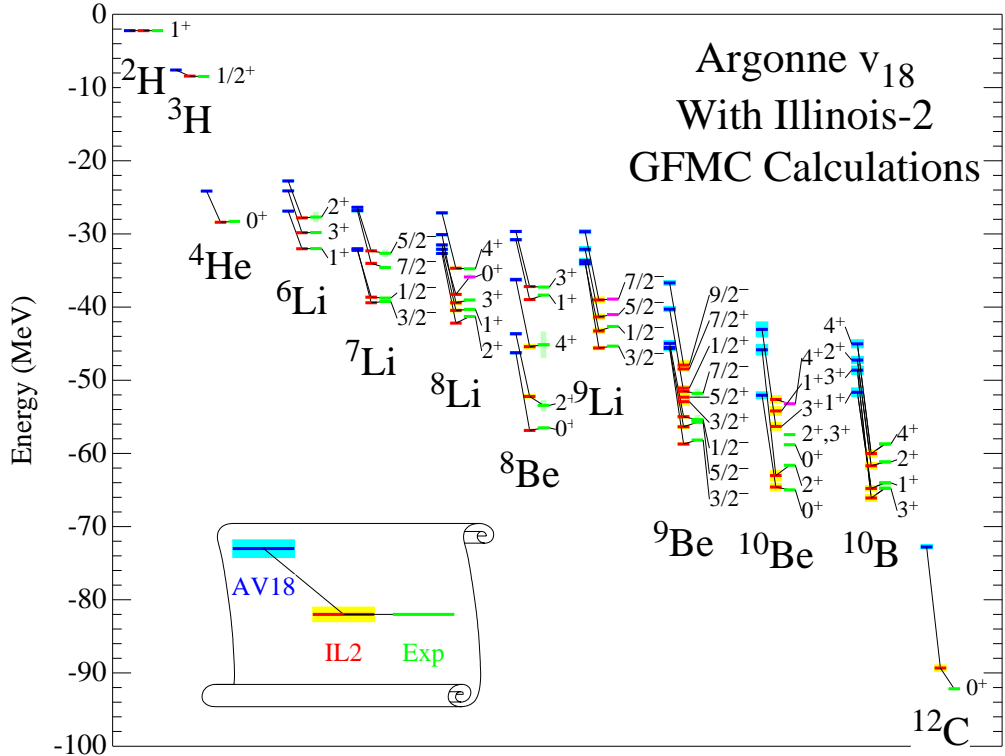}
\caption{GFMC computations of energies for the AV18 and AV18+IL2 Hamiltonians 
compared with experiment.}
\label{fig:av18-il2-exp}
\end{figure}

Figure~\ref{fig:av18-il2-exp} shows the energies of various nuclear
states.  For each isotope there are three sets of energies:  the right-most
are the experimental values, the left-most are the results of the GFMC calculations
to be described using just the \NN potential AV18, and the middle ones
are GFMC calculations using the AV18+IL2 Hamiltonian.  The AV18+IL2 results
are generally in good agreement with the data; the rms deviation is $\sim$0.75~MeV.
However without the IL2 \NNN potential the comparison to data gets
steadily worse as the number of nucleons increases.  This is a general
result that has also been obtained by others using different many-body
methods and different \NN potentials.

\begin{figure}[bt] 
\includegraphics[height=5.30in,angle=270]{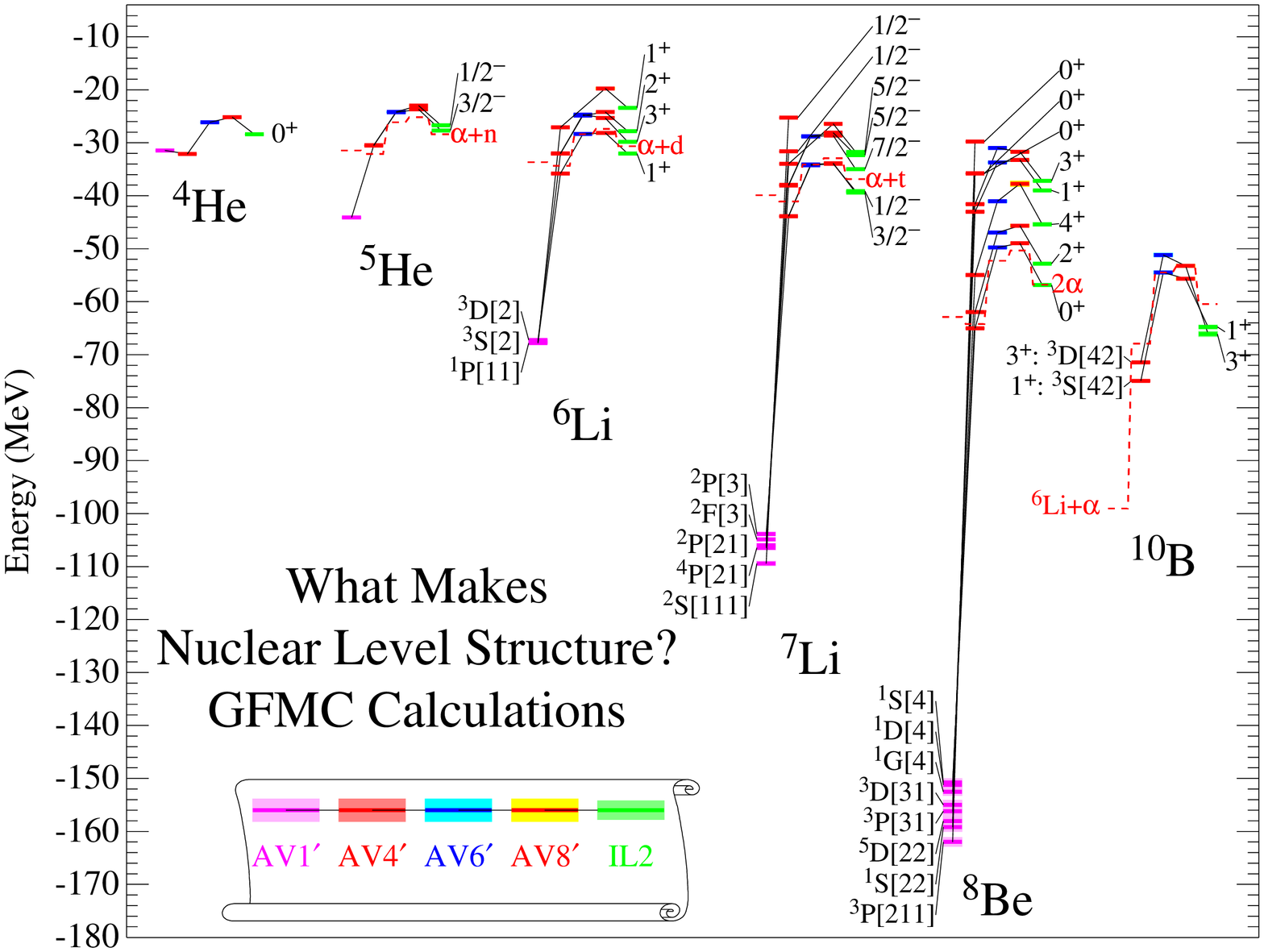}
\caption{Nuclear energy levels for various simplifications of the Hamiltonian}
\label{fig:simp-study}
\end{figure}

Figure~\ref{fig:simp-study} shows the effects of making further simplifications
to $H$ beyond removing the \NNN potential.  Here the right-most
results are again for the full AV18+IL2 Hamiltonian and thus are close
to the data.  The next set of results to left are for the AV8$^\prime$
\NN potential~\cite{PPCPW97} with no \NNN force.  With eight operators 
($[1, {\bf\sigma}_{i}\cdot{\bf\sigma}_{j}, S_{ij},
{\bf L\cdot S}]\otimes[1,{\bf\tau}_{i}\cdot{\bf\tau}_{j}]$), AV8$^\prime$
can reproduce AV18 results for eight partial waves; these
are chosen to be $^1\!S_0$, $^3\!S_1$, $^3\!D_1$, $\epsilon_1$, 
$^1\!P_1$, and $^3\!P_{0,1,2}$, ($\epsilon_1$ is the $^3\!S_1$-$^3\!D_1$
mixing angle).  (Strictly speaking, the $^3\!P_2$ potential of AV18
is reproduced but, because the $^3\!F_2$ and $\epsilon_2$ potentials are
different from those of AV18, the $^3\!P_2$ phase shifts are
not reproduced.)
This potential is more attractive in nuclei than
AV18 and more than makes up for the lost binding due to the removal
of the \NNN potential.  In general it gives a good qualitative picture
of nuclear energies.

To the left of the  AV8$^\prime$ results are results for 
AV6$^\prime$ which does not have $L{\cdot}S$ terms.  Not surprisingly,
these have only negligible spin-orbit
splittings.  In addition $^{6,7}$Li are essentially unbound to 
breakup into $\alpha$+d or $\alpha$+t 
(the dashed lines show the indicated thresholds for
each Hamiltonian). The next simplification
is AV4$^\prime$ which contains no tensor
force.  With this force, the deuteron no longer has a $D$ state
but still has the correct binding energy.  The $^1\!S_0$, $^3\!S_1$, 
$^1\!P_1$, and an average $^3\!P_J$ partial waves of AV18 are reproduced.
This simplified force results in spurious 
degeneracies of nuclear levels and
somewhat overbinds all the shown nuclei.  In particular $^8$Be
is bound against breakup into two alpha particles -- an
important failure because bound $^8$Be would result in very
different stellar nucleosynthesis.

Finally the left-most results are for AV1$^\prime$, 
a pure central force that is an average of the $^1\!S_0$ and $^3\!S_1$
potentials of AV18.  This produces an
``upside-down'' spectrum in which states with the least spatial
symmetry are most bound.  More importantly, there is no nuclear
saturation; each increase in $A$ results in much more binding
and there are no $A$=5,8 mass gaps which are essential to
big-bang nucleosynthesis and stellar evolution.

Ref.~\cite{WP02} shows results for several other Hamiltonians
including an AV2$^\prime$ that contains only central and
space-exchange terms and thus is very similar to the popular
Volkov potentials~\cite{Volkov}.  Besides erroneously binding the dineutron,
this potential has the strange feature of binding $^6$He but
not $^6$Li so that $A$=6 beta decay would be in the wrong direction.
The conclusion of this study is that one needs almost the full, 
complicated, Hamiltonian to do realistic nuclear physics.

\section{Quantum Monte Carlo Methods}
\label{sec:qmc}

The many-body problem with the full Hamiltonian described above is 
very difficult as is indicated by the slow progress over
the last half-century that is outlined
in the introduction.
We need to solve
\begin{eqnarray}
&{\cal  H}&{ \Psi(\vec{r}_1,\vec{r}_2,\cdots,\vec{r}_A;s_1,s_2,\cdots,s_A;t_1,t_2,\cdots,t_A)} \\
=&E&{ \Psi(\vec{r}_1,\vec{r}_2,\cdots,\vec{r}_A;s_1,s_2,\cdots,s_A;t_1,t_2,\cdots,t_A)} \ , \nonumber 
\end{eqnarray}
where
$s_i=\pm \case{1}{2}$ are nucleon spins, and
$t_i=\pm \case{1}{2}$ are nucleon isospins (proton or neutron).
Thus we need to solve the equivalent of
\renewcommand{\arraystretch}{.6}
$2^A\times \left(\!\!\!\begin{array}{l}A\\Z\end{array}\!\!\!\right) $ 
\renewcommand{\arraystretch}{1.2}  
complex coupled second-order equations in $3A-3$ variables 
(the number of isospin states can be reduced; see Sec.\ref{sec:representing}).
For $^{12}$C this corresponds to 270,336 coupled equations in 33 variables.

Furthermore, the coupling is strong;
the expectation value of the tensor component of $v^{\pi}$ [Eq.~\ref{eq:vpi}]
is approximately 60\% of 
the total $\langle v_{ij} \rangle$
but it is identically zero if there are no tensor correlations.
Thus we cannot perturbatively introduce the couplings.

We use two successive quantum Monte Carlo (QMC) methods to solve 
this problem.  The first is variational Monte Carlo (VMC) in which
a trial wave function, containing variational parameters, is 
posited and the expectation value of the Hamiltonian computed
using Monte Carlo integration.  In practice we have not been
able to formulate accurate enough trial wave functions and so
the second step, Green's function Monte Carlo (GFMC), is needed
to iteratively project the exact eigenfunction out of the
trial wave function.  These two methods are described in
the following sections.  School and review articles on these
methods are Refs.~\cite{CW90,P98,PW01}.   Detailed descriptions may be found
in Refs.~\cite{W91,PPCPW97,WPCP00,PVW02,PWC04}.

\section{Variational Monte Carlo}

In VMC we start with a trial wave function, $\Psi_T$, which contains 
a number of variational parameters.  We vary these parameters to
minimize the expectation value of $H$, 
\begin{eqnarray}
E_T = \frac{\langle \Psi_T | H | \Psi_T \rangle}
{\langle \Psi_T | \Psi_T \rangle} \geq E_0  \ .
\end{eqnarray}
As indicated, the resulting $E_T$ is, by the Raleigh-Ritz variational
principle, greater than the true ground-state energy for the quantum
numbers ($J^\pi$, $J_z$, $T$, and $T_z$) of $\Psi_T$.  A 
simplified form of our trial wave functions is
\begin{eqnarray}
| \Psi_T \rangle = 
[{\cal S}\prod_{i<j}(1+U_{ij} + \Sigma_{k} U_{ijk})] 
\prod_{i<j}f_c(r_{ij}) | { \Phi} \rangle \ .
\label{psit}
\end{eqnarray}
Here 
$f_c(r)$ is a  central (mostly short-ranged repulsion) correlation,
$U_{ij}$ are non-commuting two-body correlations induced by $v_{ij}$, and
$U_{ijk}$ is a simplified three-body correlation from $V_{ijk}$.

More specifically, 
\begin{eqnarray}
U_{ij} = \sum_{p=2,6} u_{p}(r_{ij}) O^{p}_{ij} \ ,
\end{eqnarray}
contains 
${\bf\tau}_{i}\cdot{\bf\tau}_{j}$, ${\bf\sigma}_{i}\cdot{\bf\sigma}_{j}$, 
${\bf\sigma}_{i}\cdot{\bf\sigma}_{j} \ {\bf\tau}_{i}\cdot{\bf\tau}_{j}$, 
$S_{ij}$, and $S_{ij} \ {\bf\tau}_{i}\cdot{\bf\tau}$
operators, of which the $S_{ij} \ {\bf\tau}_{i}\cdot{\bf\tau}$ is most important
due to the already noted strong tensor contribution from $v^{\pi}$.
The $f_c(r)$ and $u_{p}(r)$ are solutions of coupled differential equations 
with $v_{ij}$ as input~\cite{W91}.

The $\Phi$ (see below) is fully antisymmetric; hence the rest of Eq.~(\ref{psit})
must be symmetric.  
But the $U_{ij}$ do not commute; for example
\begin{equation}
[ \sigma_1 \cdot \sigma_2 \,, \sigma_1 \cdot \sigma_3 ] =
2i \, \sigma_1 \cdot ( \sigma_2 \times \sigma_3 ) \ .
\end{equation}
The symmetrizer $\cal S$ fixes this by summing over all
$[{A(A-1)\over 2}]!$ permutations of the ordering in $\prod_{i<j}$.
In practice this is done by using just one Monte Carlo chosen ordering per 
wave function evaluation.

\subsection{The one-body part of $\Psi_T$, $\Phi$}

The one-body part of $\Psi_T$, $\Phi$, is a $1 \hbar\omega$ shell-model wave function.
It determines the quantum numbers of the state being computed and is
fully antisymmetric.
For $^3$H and $^{3,4}$He, $\Phi$ can be antisymmetrized in just spin-isospin space,
for example
\begin{eqnarray}
|\Phi (^3H, M_J=\hbox{$\frac{1}{2}$}) \rangle = \hbox{$\frac{1}{\sqrt{6}}$} (
|p\!\uparrow n\!\uparrow n\!\downarrow \rangle &-& | p\!\uparrow n\!\downarrow n\!\uparrow \rangle
+ |n\!\downarrow p\!\uparrow n\!\uparrow \rangle \\
 ~~- |n\!\uparrow p \!\uparrow n\!\downarrow \rangle
&+& |n\!\uparrow n\!\downarrow p\!\uparrow \rangle - |n\!\downarrow n\!\uparrow p\!\uparrow \rangle
\, ) \ . \nonumber
\end{eqnarray}

For $A > 4$ we need $P$-wave radial wave functions in order to antisymmetrize $\Phi$; the 
antisymmetrization is achieved by summing over all partitions of the $A$
nucleons into four $S$-shell nucleons (the $\alpha$ core) and $A$-4 $P$-shell
nucleons which are antisymmetrically coupled to $J^\pi$ and $T$.
To make $\Phi$ translationally invariant, we express all functions of single-particle positions
as functions of position relative to the center of mass of the $A$ nucleons
or of some sub-cluster of them.
The one-body wave functions are solutions of Woods Saxon potentials containing several
variational parameters.  If desired the separation energy of the these
one-body wave functions can be fixed at the experimental value to guarantee that
the $\Psi_T$ has the correct asymptotic form.
In general $\Phi$ has several spatial-symmetry components depending on
how many ways a state of the desired quantum numbers can be constructed in 
the $P$-shell basis.  For example the $^6$Li $\Phi$ have the form
\begin{eqnarray}
 &&  |\Phi\rangle =  {\cal A}  \sum_{LS} \beta_{LS} 
        |\Phi_6(LSJMTT_{3})_{1234:56}\rangle  \ ,  \\
 &&  \Phi_{6}(LSJMTT_{3})_{1234:56} = 
        \Phi_{4}(0 0 0 0)_{1234} 
        \phi^{LS}_{p}(R_{\alpha 5}) \phi^{LS}_{p}(R_{\alpha 6})   \\
 && ~~~~ \left\{ [Y_{1m_l}(\Omega_{\alpha 5}) Y_{1m_l'}(\Omega_{\alpha 6})]_{LM_L}
     \times [\chi_{5}(\frac{1}{2}m_s) \chi_{6}(\frac{1}{2}m_s')]_{SM_S}
     \right\}_{JM}     \nonumber \\
  && ~~~~ \times [\nu_{5}(\frac{1}{2}t_3) \nu_{6}(\frac{1}{2}t_3')]_{TT_3} 
   \ .  \nonumber 
\end{eqnarray}
A $1 \hbar\omega$ $LS$-basis diagonalization determines the $\beta_{LS}$.

\subsection{Representing $\Psi_T$ in the computer}
\label{sec:representing}

The wave function, $\Psi_T( \vec r_1 , \vec r_2 , \cdots , \vec r_A )$, 
is a complex vector in spin-isospin space with dimension
$[N_S$~components~for~spin$] \times [N_T$~components~for~isospin$]$.
The number of spin states is $2^A$.  However for even $A$, if
we choose to use $M_J=0$, we can calculate and retain only
half the spin vector, say that part with positive spin for the
last nucleon, and obtain the other half of the vector by
time-reversal symmetry.
The number of isospin states, $N_T$, depends on the isospin basis being used:
\renewcommand{\arraystretch}{.8}
\begin{eqnarray}
N_T &=& \left( \!\! \begin{array}{c} A \\ Z \end{array} \!\! \right) 
 \ \ \ \ \ \ \ \ \ \ \ \  \ \ \ \ \ \ \ \ \ \ \ \ \ \ \ \ \ \ \ 
\mbox{for a proton-neutron basis} \ ,\\
    &=& {2T+1\over A/2+T+1} \, \left( \!\! \begin{array}{c} A \\A/2+T \end{array} \!\! \right) 
~~~ \mbox{ for a good isospin basis} \ .
\end{eqnarray}
\renewcommand{\arraystretch}{1.2}  

Potentials ($v_{ij}$, $V_{ijk}$) and correlations ($u_{ij}$, $U_{ijk}$) involve
repeated operations on $\Psi$.  For example $\sigma_i \cdot \sigma_j$ may
be written as
\renewcommand{\arraystretch}{1.0}
\begin{eqnarray}
\sigma_i \cdot \sigma_j &=& 2 ( \sigma^+_i \sigma^-_j + \sigma^-_i
\sigma^+_j ) +  \sigma^z_i \sigma^z_j \\
&=& 2P^{\sigma}_{ij} - 1 \\
&=& \left(
\begin{array}{crrc}
1 & 0 & 0 & 0 \\
0 & -1 & 2 & 0 \\
0 & 2 & -1 & 0 \\
0 & 0 & 0 & 1
\end{array} \right) 
 \ \mbox{acting on} \ \left(
\begin{array}{c}
\uparrow\uparrow     \\
\uparrow\downarrow   \\
\downarrow\uparrow   \\
\downarrow\downarrow 
\end{array} \right) \ .
\end{eqnarray}
Here $P^{\sigma}_{ij}$ exchanges the spin of $i$ and $j$. 
Consider the spin part of an $A$=3 wave function; 
$\sigma_i \cdot \sigma_j$ will not mix different isospin components
and, for different $i$ and $j$, will separately act on different, 
non-contiguous, 4-element blocks of $\Psi$:
\begin{eqnarray}
\Psi =
\left(
\begin{array}{c}
a_{\uparrow\uparrow\uparrow}\\a_{\uparrow\uparrow\downarrow}\\
a_{\uparrow\downarrow\uparrow}\\a_{\uparrow\downarrow\downarrow}\\
a_{\downarrow\uparrow\uparrow}\\a_{\downarrow\uparrow\downarrow}\\
a_{\downarrow\downarrow\uparrow}\\a_{\downarrow\downarrow\downarrow}
\end{array} \right) &;& \;\;
\sigma_1 \cdot \sigma_2 \Psi = \left(
\begin{array}{c}
 a_{\uparrow\uparrow\uparrow}\\
 a_{\uparrow\uparrow\downarrow}\\
 2a_{\downarrow\uparrow\uparrow}-a_{\uparrow\downarrow\uparrow}\\
 2a_{\downarrow\uparrow\downarrow}-a_{\uparrow\downarrow\downarrow}\\
 2a_{\uparrow\downarrow\uparrow}-a_{\downarrow\uparrow\uparrow}\\
 2a_{\uparrow\downarrow\downarrow}-a_{\downarrow\uparrow\downarrow}\\
 a_{\downarrow\downarrow\uparrow}\\
 a_{\downarrow\downarrow\downarrow}
\end{array} \right) ; \label{eq:sigsig1} \\
\sigma _2 \cdot \sigma _3 \Psi = \left(
\begin{array}{c}
 a_{\uparrow\uparrow\uparrow}\\
 2a_{\uparrow\downarrow\uparrow}-a_{\uparrow\uparrow\downarrow}\\
 2a_{\uparrow\uparrow\downarrow}-a_{\uparrow\downarrow\uparrow}\\
 a_{\uparrow\downarrow\downarrow}\\
 a_{\downarrow\uparrow\uparrow}\\
 2a_{\downarrow\downarrow\uparrow}-a_{\downarrow\uparrow\downarrow}\\
 2a_{\downarrow\uparrow\downarrow}-a_{\downarrow\downarrow\uparrow}\\
 a_{\downarrow\downarrow\downarrow}
\end{array} \right) &;& \;\;
\sigma _3 \cdot \sigma _1 \Psi = \left(
\begin{array}{c}
 a_{\uparrow\uparrow\uparrow}\\
 2a_{\downarrow\uparrow\uparrow}-a_{\uparrow\uparrow\downarrow}\\
 a_{\uparrow\downarrow\uparrow}\\
 2a_{\downarrow\downarrow\uparrow}-a_{\uparrow\downarrow\downarrow}\\
 2a_{\uparrow\uparrow\downarrow}-a_{\downarrow\uparrow\uparrow}\\
 a_{\downarrow\uparrow\downarrow}\\
 2a_{\uparrow\downarrow\downarrow}-a_{\downarrow\downarrow\uparrow}\\
 a_{\downarrow\downarrow\downarrow}
\end{array} \right)  \ .
\label{eq:sigsig2}
\end{eqnarray}
\renewcommand{\arraystretch}{1.2}  
Similarly, the tensor operator is
\begin{eqnarray}
S_{ij} &=& 3 \,\sigma _i \cdot \hat r_{ij} \,\sigma _j \cdot \hat r_{ij} -
\sigma _i \cdot \sigma _j \\
&=& 3 \left(
\begin{array}{llllllll}
z^2 - 1/3 && z(x-iy) && z(x-iy) && (x-iy)^2\\
z(x+iy) && -z^2 - 1/3 && x^2 + y^2 - 2/3 && -z(x-iy)\\
z(x_iy) && x^2+y^2-2/3 && -z^2+1/3 && -z(x-iy)\\
(x+iy)^2 && -z(x+iy) && -z(x+iy) && z^2-1/3
\end{array} \right) \ ,
\end{eqnarray}
where $x=x_i-x_j$, etc.
As shown in Eqs.~(\ref{eq:sigsig1}) and~(\ref{eq:sigsig2}), these $4\times4$
matrices form a sparse matrix of (non-contiguous) $4\times4$ blocks
in the $A$-body problem.  Specially coded subroutines are used
to efficiently perform these operations.

Most of the time in VMC or GFMC calculations is spent evaluating
wave functions (or in GFMC making a propagation step which is equivalent).
The pair operators dominate this time.  The evaluation of a
kinetic energy involves numerical second derivatives which require $6A$ 
wave function computations.  Hence the product of $A$, the number of 
pairs, and of the length of the spin-isospin vector is a good indication
of how the total computational time scales with $A$.  
Table~\ref{tab:psi_time} shows this scaling for various nuclei,
assuming $M$=0 for even $J$ nuclei and that good-isospin bases
are being used.  The final column shows the product of the first
three columns relative to $^8$Be.
We can do calculations up to $A$=10 routinely and
a few $^{12}$C calculations have been done.  It is clear that
this approach is not reasonable for $^{16}$O.  The last two lines
are for ``neutron drops'' for which isospin does not have to
be considered.  This allows somewhat bigger $A$ to be reached.

\begin{table}[bt]
\centering
\begin{minipage}[t]{3.50in}
\begin{tabular}{rrrrr}
\hline                                                        
           & $A$ & Pairs &
                            Spin$\times$Isospin & $\prod (/^8$Be) \\
\hline
   $^4$He  &  4  &   6~~~ &     8$\times$2~~~~~~~~   &   0.001~~~     \\
   $^5$He  &  5  &  10~~~ &    32$\times$5~~~~~~~~   &   0.020~~~     \\
   $^6$Li  &  6  &  15~~~ &    32$\times$5~~~~~~~~   &   0.036~~~     \\
   $^7$Li  &  7  &  21~~~ &   128$\times$14~~~~~~~   &   0.66~~~~     \\
   $^8$Be  &  8  &  28~~~ &   128$\times$14~~~~~~~   &   1.~~~~~~~     \\
   $^8$Li  &  8  &  28~~~ &   128$\times$28~~~~~~~   &   2.~~~~~~~     \\
   $^9$Be  &  9  &  36~~~ &   512$\times$42~~~~~~~   &  18.~~~~~~~     \\
 $^{10}$B  & 10  &  45~~~ &   512$\times$42~~~~~~~   &  24.~~~~~~~     \\
 $^{10}$Be & 10  &  45~~~ &   512$\times$90~~~~~~~   &  51.~~~~~~~     \\
\hline                                                           
 $^{11}$B  & 11  &  55~~~ &  2048$\times$132~~~~~~   & 400.~~~~~~~     \\
 $^{12}$C  & 12  &  66~~~ &  2048$\times$132~~~~~~   & 530.~~~~~~~     \\
\hline                                                           
 $^{16}$O  & 16  & 120~~~ & 32768$\times$1430~~~~~   & 224,000.~~~~~~~  \\
 $^{40}$Ca & 40  & 780~~~ & 3.6$\times 10^{21}\times 6.6$$\times 10^9$ & 2.8$\times 10^{20}$  \\
\hline                                                           
   $^8$n   &  8  &  28~~~ &   128$\times$1~~~~~~~~~  &   0.071~~~     \\
 $^{14}$n  & 14  &  91~~~ &  8192$\times$1~~~~~~~~~  &  26.~~~~~~~    \\
\hline                                                        
\end{tabular}
\caption{Scaling of wave function computation time}
\label{tab:psi_time}
\end{minipage}
\end{table}

\subsection{A Variational Monte Carlo Calculation}

The basic steps in a variational calculation are

\begin{itemize}
\renewcommand{\itemsep}{-4pt}
\item Generate a random position:  ${\bf R} = \vec r_1 , \vec r_2 , \cdots , \vec r_A$ .
\item Make many (1000's) random steps based on the probability  $P = |\Psi_T({\bf R})|^2$ .
\item Start integration loop:  \\
-- Make order 10 steps based on $P$ . \\
-- Compute and sum 
$H_{\mbox{local}}({\bf R}) = [\Psi_T({\bf R})^\dagger H \Psi_T({\bf R})] / |\Psi_T({\bf R})|^2$ .
Gradients and Laplacians are computed by differences:  
$6A$ evaluations of $\Psi_T({\bf R} + \delta_j \vec r_i)$ .
\item $\langle \Psi_T | H | \Psi_T \rangle /  \langle \Psi_T | \Psi_T \rangle =$ average$(H_{\mbox{local}})$ .
\end{itemize}

A random step from a given position, ${\bf R}$, to a
new position ${\bf R}^{\prime}$ is made using the Metropolis method:

\begin{itemize}
\renewcommand{\itemsep}{-4pt}
\item Use $3A$ uniform random numbers on (0,1), $\{w_j\}$,  
to make $\triangle{\bf R} ; ~~ \triangle x_i = 2 \delta r (w_j - 1)$ .
\item Set ${\bf R}^\prime = {\bf R} + \triangle{\bf R}$, and compute
$P({\bf \triangle R}) = |\Psi_T({\bf R^\prime})|^2 / |\Psi_T({\bf R})|^2$ .
\item Make another  random number on (0,1):  $p$
\item If $P > p$, the step is accepted; replace ${\bf R}$ with ${\bf R'}$ . \\
      if $P < p$, the step is rejected; discard ${\bf R'}$ and stay at ${\bf R}$ .
\end{itemize}

\subsection{Accuracy of VMC energies}

\begin{figure}[bt] 
\centering
\includegraphics[height=5.0in,angle=270]{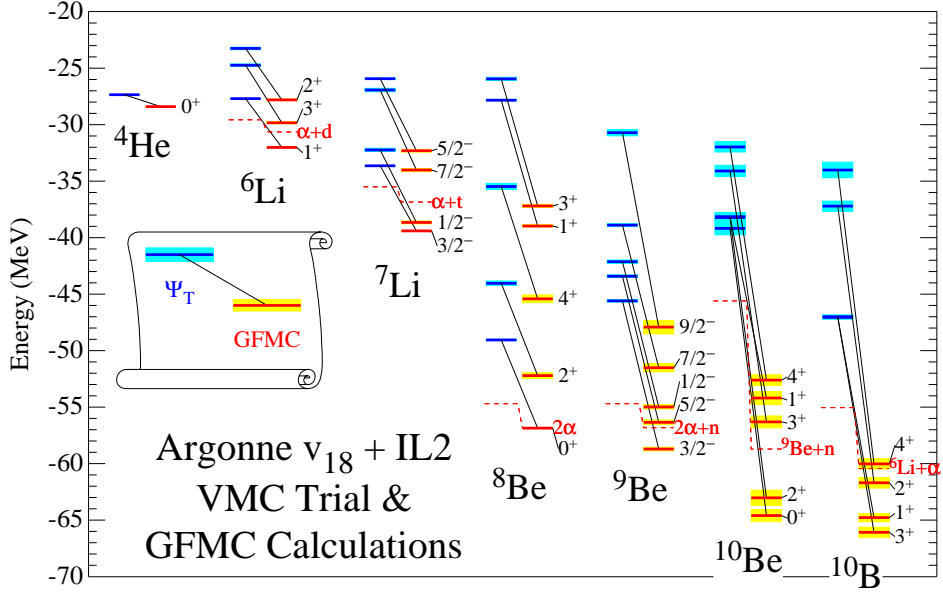}
\caption{Comparison of VMC and GFMC energies for the AV18+IL2 Hamiltonian.
The light shading shows Monte Carlo statistical errors.}
\label{fig:vmc-gfmc}
\end{figure}

Figure \ref{fig:vmc-gfmc} compares VMC energies of various nuclear
states with the corresponding GFMC values for the AV18+IL2 Hamiltonian.
As is described in the next section, the GFMC results are believed to be
accurate to 1--2\%.  For $^4$He the VMC result is quite close to the
GFMC.  However as we move into the $P$~shell, the VMC results get
steadily worse.  In fact, although the GFMC calculations show that this
Hamiltonian binds the nuclei shown with the exception of $^8$Be, the VMC
energies are all above the VMC energies for the subclusters that the
nuclei can breakup into.  Furthermore the $^8$Be VMC energy is actually
lower than those of 9- and 10-body nuclei.  Calculations with simpler
Hamiltonians show that these failures of the VMC energies are related to
the tensor force; VMC calculations for the simple AV4$^{\prime}$
potential discussed in Sec.~\ref{sec:avx} are quite accurate, while
those for AV6$^{\prime}$ have significant errors.

\section{Green's Function Monte Carlo -- General Description}
\label{sec:gfmc}

As shown above, our VMC trial wave functions are not good enough
for $P$-shell nuclei.  This means that they contain admixtures
of excited-state components in addition to the desired exact ground-state
component, $\Psi_0$;
\begin{eqnarray}
  \Psi_T = \Psi_0 + {\sum}\alpha_i\Psi_i  \ .
\end{eqnarray}
We use Green's Function Monte Carlo 
to project $\Psi_0$ out of $\Psi_T$ by propagating in imaginary time, $\tau$:
\begin{eqnarray}
\Psi(\tau) &=& \exp [ - ( H - \tilde{E}_0) \tau ] \Psi_T \ ,  \\
 &=& e^{-(E_0-\tilde{E}_0)\tau} \times 
         [ \Psi_0 + {\sum}\alpha_i e^{-(E_i-E_0)\tau} \Psi_i ] \ , \\
\lim_{\tau \rightarrow \infty} \Psi(\tau) &\propto& \Psi_0 \ ,
\end{eqnarray}
where $\tilde{E}_0$ is a guess for the exact $E_0$.

The eigenvalue $E_{0}$ is calculated exactly while other expectation values
are generally calculated neglecting terms of order $|\Psi_{0}-\Psi_{T}|^{2}$ 
and higher.  
In contrast, the error in the variational energy, $E_{T}$, is of order 
$|\Psi_{0}-\Psi_{T}|^{2}$, and other expectation values calculated with 
$\Psi_{T}$ have errors of order $|\Psi_{0}-\Psi_{T}|$.

The evaluation of $\Psi(\tau)$ is made by
introducing a small time step, $\triangle\tau$, $\tau=n\triangle\tau$, 
\begin{eqnarray}
\Psi(\tau) = \left[e^{-({H}-E_{0})\triangle\tau}\right]^{n} \Psi_{T}
= G^n \Psi_{T} \ .
\end{eqnarray}
where $G$ is the short-time Green's function.
The $\Psi (\tau)$ is represented by a vector function of $\bf R$, and the
Green's function, $G_{\alpha\beta}({\bf R}^{\prime},{\bf R})$ is a matrix
function of $\bf R^{\prime}$ and ${\bf R}$ 
in spin-isospin space, defined as
\begin{eqnarray}
G_{\alpha\beta}({\bf R}^{\prime},{\bf R})= \langle {\bf 
R}^{\prime},\alpha|e^{-({H}-E_{0})\triangle\tau}|{\bf R},\beta\rangle \ .
\label{eq:gfunction}
\end{eqnarray}
It is calculated with leading errors of order $(\triangle\tau)^{3}$ as
discussed below.  Omitting
spin-isospin indices for brevity, $\Psi({\bf R}_{n},\tau)$ is given by
\begin{eqnarray}
\Psi({\bf R}_{n},\tau) = \int G({\bf R}_{n},{\bf R}_{n-1})\cdots G({\bf 
R}_{1},{\bf R}_{0})\Psi_{T}({\bf R}_{0}) \ d{\bf P} \ ,
\label{eq:gfmcpsi}
\end{eqnarray}
and
\begin{eqnarray}
E(\tau) = \frac{ \int  \Psi_{T}^\dagger({\bf R}_{n}) \
G^\dagger({\bf R}_{n},{\bf R}_{n-1})\cdots G^\dagger({\bf R}_{1},{\bf R}_{0})
 \ H \ \Psi_{T}({\bf R}_{0}) \ d{\bf P} }
{\int  \Psi_{T}^\dagger({\bf R}_{n}) \
G^\dagger({\bf R}_{n},{\bf R}_{n-1})\cdots G^\dagger({\bf R}_{1},{\bf R}_{0})
  \ \Psi_{T}({\bf R}_{0}) \ d{\bf P} } \ ,
\label{eq:gfmcE}
\end{eqnarray}
where $d{\bf P} = d{\bf R}_{0} d{\bf R}_{1}\cdots d{\bf R}_{n}$.
Here we have placed the $\Psi(\tau)$ to the left side of $H$
because the derivatives in $H$ may be evaluated only on $\Psi_T$; 
we cannot compute gradients or Laplacians of $\Psi(\tau)$.
This $3An$-dimensional integral is computed by Monte Carlo.

\subsection{The Short-Time Propagator}

The success of a GFMC calculation depends on an accurate and fast
evaluation of the short-time propagator, 
$G_{\alpha\beta}({\bf R}^{\prime},{\bf R})$.  One wants to be
able to do this for the largest possible value of $\triangle\tau$
to reduce the number of steps, $n$, needed to reach some asymptotic
value of $\tau$.  The most important features of $\Psi(\tau)$ are
induced by the \NN potential, so consider first 
$G_{\alpha\beta}({\bf R}^{\prime},{\bf R})$ for a Hamiltonian
with no \NNN potential.  This can be written as
\begin{eqnarray}
G_{\alpha\beta}({\bf R}^{\prime},{\bf R})= e^{\tilde{E}_0\triangle\tau}
G_{0}({\bf R^{\prime}},{\bf R})
\langle\alpha|\left[{\cal S}\prod_{i<j}\frac{g_{ij}
({\bf r}_{ij}^{\prime},{\bf r}_{ij})}{g_{0,ij}({\bf r}^{\prime}_{ij},{\bf r}_{ij})} 
\right] |\beta\rangle \ ,
\label{eq:propagator2}
\end{eqnarray}
where
\begin{eqnarray}
G_{0}({\bf R}^{\prime},{\bf R}) = \langle {\bf R}^{\prime}|e^{-{K}\triangle\tau}|{\bf 
R}\rangle =  \left[ \sqrt{\frac{m}
{2\pi\hbar^{2} \triangle\tau}}\, \right]^{3A}\exp\left[\frac{-({\bf R}^{\prime}-{\bf 
R})^2}{2\hbar^{2}\triangle\tau/m}\right] \ ,
\label{eq:gaussian}
\end{eqnarray}
is the many-nucleon free propagator and $g_{0,ij}$ is the corresponding
two-nucleon free propagator,
\begin{eqnarray}
g_{0,ij}({\bf r}_{ij}^{\prime},{\bf r}_{ij}) = \left[ 
\sqrt{\frac{\mu}{2\pi\hbar^{2} \triangle\tau}}\, 
\right]^3 \exp\left[-\frac{({\bf r}_{ij}^{\prime}-{\bf r}_{ij})^{2}}
{2\hbar^{2}\triangle\tau/\mu}\right] \ ,
\end{eqnarray}
and $\mu = m/2$ is the reduced mass. 

The $G_{0}({\bf R}^{\prime},{\bf R})$ is included in the Monte Carlo
integration [Eq.~(\ref{eq:gfmcE})] by using it to make the step
from $\bf R$ to ${\bf R}^{\prime}$.  The magnitudes of the $3A$ steps
($x$, $y$, and $z$ for each nucleon) are determined by sampling
a Gaussian of the width given in Eq.~(\ref{eq:gaussian}) and the directions
of the steps are picked by importance sampling; see Ref.~\cite{PPCPW97}
for details.

Eq.~(\ref{eq:propagator2}) introduces the exact two-body propagator,
\begin{eqnarray}
g_{ij}({\bf r}_{ij}^{\prime},{\bf r}_{ij}) &=& \langle{\bf r}_{ij}^{\prime}
|e^{-H_{ij}\triangle\tau}|{\bf r}_{ij}\rangle~, \\
\label{eq:exactg}
{H}_{ij} &=& -\frac{\hbar^{2}}{m}\nabla^{2}_{ij} + v_{ij} \ .
\end{eqnarray}
All terms containing any
number of the same $v_{ij}$ and $K$ are treated exactly in this propagator, 
as we have included the imaginary-time equivalent of the full two-body
scattering amplitude.
Eq.~(\ref{eq:propagator2}) still has errors of order $(\triangle\tau)^{3}$, 
however they are from
commutators of 
terms like $v_{ij}Kv_{ik}(\triangle\tau)^{3}$ which become large only when
both pairs $ij$ and $ik$ are close.

To calculate $g_{ij}$, we use the techniques developed by Schmidt and
Lee\cite{SL95} for scalar interactions.  These allow $g_{ij}$ to
be calculated with high ($\sim$ 10 digit) accuracy.
However, this calculation is quite time consuming.  Therefore,
prior to the GFMC calculation, we compute
and store the the propagator on a grid.  
For a spin-independent interaction, the propagator $g_{ij}$ would depend
only upon the two magnitudes $r^{\prime}$ and $r$ and the angle
$cos (\theta) =  \hat{\bf r}^{\prime} \cdot \hat{\bf r}$ between
them.  
Here, though, there is also a dependence upon the spin quantization axis.
Rotational symmetry allows one 
to calculate the spin-isospin components of
$g_{ij}({\bf r}^{\prime},{\bf r})$ for 
any ${\bf r}^{\prime}$ and ${\bf r}$ by simple SU3 
spin rotations and values of $g_{ij}$
on a grid of initial points ${\bf r}=(0,0,z)$ and final points 
${\bf r}^{\prime}=(x^{\prime},0,z^{\prime})$.  
In addition, the fact that
the propagator is Hermitian allows us to store only the values for
$z > z^{\prime}$.

Returning to the full Hamiltonian including \NNN forces, 
the complete propagator is given by
\begin{eqnarray}
G_{\alpha\beta}({\bf R}^{\prime},{\bf R}) &=& e^{E_0\triangle\tau}
G_{0}({\bf R}^{\prime},{\bf R})\exp[{- \sum (V^{R}_{ijk}({\bf R}^{\prime})
+ V^{R}_{ijk}({\bf R}))\frac{\triangle\tau}{2}}] \nonumber \\ 
& & \langle\alpha|I_{3}({\bf R}^{\prime})|\gamma\rangle\langle\gamma|\left[{\cal 
S}\prod_{i<j}\frac{g_{ij}({\bf r}_{ij}^{\prime},{\bf r}_{ij})}{g_{0,ij}({\bf 
r}_{ij}^{\prime},{\bf r}_{ij})} \right] |\delta\rangle\langle\delta|I_{3}({\bf 
R})|\beta\rangle~,
\label{eq:fullprop}
\end{eqnarray}
with
\begin{eqnarray}
I_{3}({\bf R}) = \left[1 - \frac{\triangle\tau}{2}\sum V^{2\pi}_{ijk}({\bf 
R})\right]~.
\end{eqnarray}
The exponential of $V^{2\pi}_{ijk}$ is expanded to first order in 
$\triangle\tau$ thus, there are additional error terms of the form 
$V^{2\pi}_{ijk}V^{2\pi}_{i'j'k'}(\triangle\tau)^{2}$.  However, they have 
negligible effect since $V^{2\pi}_{ijk}$ has a magnitude of only a few MeV.
It was verified that the results for $^{4}$He do not show any change, outside
of statistical errors, when $\triangle\tau$ is decreased from 0.5 GeV$^{-1}$.

\subsection{Problems with Nuclear GFMC}

While GFMC is in principal exact for the $\langle H \rangle$, there
are several practical difficulties that make it only approximate.
We have made many tests of the accuracy of the GFMC energies, both
by comparison to other methods and by comparing calculations with
different $\triangle \tau$, starting $\Psi_T$, and other computational
parameters.  These tests show that our results for energy are good
to $\sim$1\% up to $\sim$2\% for larger $A$ or $N-Z$ ($^8$He is particularly
difficult).  Some of the problems are:

\subsubsection{Limitation on $H$}
The exact propagator of Eq.~(\ref{eq:exactg}) can be computed for the 
full $v_{18}$ potential, however the ${\bf L}^2$ and $({\bf L} \cdot {\bf S})^2$
terms in the potential correspond to state-dependent changes of the
mass appearing in the free Green's function.  Since we do not know
how to sample such a free Green's function, we cannot use the
exact $g_{ij}$ for the full potential, but rather must use one
constructed for an approximately equivalent potential that does not
contain quadratic ${\bf L}$ terms, namely the AV8$^\prime$
introduced in Sec.~\ref{sec:avx}.  The difference between the desired
and approximate potentials is computed perturbatively.  Comparisons
with more accurate calculations for $^3$H and $^4$He suggest that
this introduces errors of less than 1\%.

\subsubsection{Fermion sign problem}

The $G({\bf R}_{i},{\bf R}_{i-1})$ is a local operator and can mix in
the boson solution.  This has a (much) lower energy than the fermion
solution and thus is exponentially amplified in subsequent propagations.
In the final integration with the antisymmetric $\Psi_T$, the desired
fermionic part is projected out in Eq.~(\ref{eq:gfmcE}), but in the
presence of large statistical errors that grow exponentially with
$\tau$.  Because the number of pairs that can be exchanged grows with
$A$, the sign problem also grows exponentially with increasing $A$.  For
$A{\geq}8$, the errors grow so fast that convergence in $\tau$ cannot be
achieved.

For simple scalar wave functions, the fermion sign problem can be controlled
by not allowing the propagation to move across a node of the wave function.
Such ``fixed-node'' GFMC provides an approximate solution which is the
best possible variational wave function with the same nodal structure 
as $\Psi_T$.
However, a more complicated
solution is necessary for the spin- and isospin-dependent wave functions
of nuclei.  This is provided by ``constrained-path'' propagation in which
those 
configurations that, in future generations, will contribute only noise to 
expectation values are discarded.
If the exact ground state $| \Psi_0 \rangle$ were known,
any configuration at time step $n$ for which
\begin{equation}
\Psi({\bf R}_n)^\dagger \Psi_0({\bf R}_n) = 0 \ ,
\label{eq:gfmc:const_config}
\end{equation}
where a sum over spin-isospin states is implied, could be discarded.
The sum of these discarded
configurations can be written as a state $| \Psi_d \rangle$,
which obviously has zero overlap with the ground state.
The $\Psi_d$ contains only excited states and should decay away as 
$\tau \rightarrow \infty$, thus discarding it is justified. 
Of course the exact $\Psi_0$ is not known, and so configurations
are discarded with a probability such that the average overlap
with the trial wave function, 
\begin{equation}
\langle \Psi_d | \Psi_T \rangle = 0 \ .
\end{equation}

Many tests of this procedure have been made~\cite{WPCP00} and it usually gives
results that are consistent with unconstrained propagation, within statistical errors.
However a few cases in which the constrained propagation converges to the
wrong energy (either above or below the correct energy) have been found.
Therefore a small number, $n_u=10$ to 20, of unconstrained steps are made
before evaluating expectation values.  These few unconstrained steps,
out of typically 400 total steps,
appear to be enough to damp out errors introduced by the constraint,
but do not greatly increase the statistical error.  Unfortunately,
the constrained-path $E(\tau)$ are not upper bounds to the true
$E_0$; examples have been found in which the constrained energies
evaluated with inadequate $n_u$ are below $E_0$.

\subsubsection{Mixed estimates extrapolation}
As shown in Eq.~(\ref{eq:gfmcE}), GFMC computes ``mixed'' expectation values
between $\Psi_T$ and $\Psi(\tau)$ of operators,
\begin{eqnarray}
\langle O \rangle_{\rm Mixed} & = & \frac{\langle \Psi(\tau) | O |
\Psi_{T}\rangle}{\langle \Psi(\tau) | \Psi_{T}\rangle} \ .
\end{eqnarray}
The desired expectation values, of course, have $\Psi(\tau)$ on both
sides. By writing
$\Psi(\tau) = \Psi_{T} + \delta\Psi(\tau)$ 
and neglecting terms of order $[\delta\Psi(\tau)]^2$, we obtain the approximate
expression
\begin{eqnarray}
\langle O (\tau)\rangle =
\frac{\langle\Psi(\tau)| O |\Psi(\tau)\rangle}
{\langle\Psi(\tau)|\Psi(\tau)\rangle}
\approx \langle O (\tau)\rangle_{\rm Mixed}
     + [\langle O (\tau)\rangle_{\rm Mixed} - \langle O \rangle_T] ~,
\label{eq:pc_gfmc}
\end{eqnarray}
where $\langle O \rangle_T$ is the variational expectation value.
More accurate evaluations of $\langle O (\tau)\rangle$ are possible,
essentially by measuring the observable at the mid-point of the path.
However, such estimates require a propagation twice as long as the mixed
estimate and require separate propagations for every $\langle O \rangle$
to be evaluated.

The expectation value of the Hamiltonian is a special case.
The $\langle{H}(\tau)\rangle_{\rm Mixed}$ can be re-expressed as~\cite{CK79}
\begin{eqnarray}
\langle{H}(\tau)\rangle_{\rm Mixed} = \frac{\langle \Psi_{T} |
e^{-({H}-E_{0})\tau /2}{H} e^{-({H}-E_{0})\tau /2} |
\Psi_{T}\rangle}{\langle \Psi_{T} |e^{-({H}-E_{0})\tau /2}
e^{-({H}-E_{0})\tau /2}| \Psi_{T}\rangle} \geq E_{0}~,
\end{eqnarray}
since the propagator $\exp [ - (H - E_0) \tau ] $ commutes with the
Hamiltonian.
Thus $\langle{H}(\tau)\rangle_{\rm Mixed}$ is already the correct
expectation value and must not be extrapolated.  This results
in the unfortunate circumstance that the sum of the pieces of
$\langle H \rangle$ is not equal to the full GFMC value
of $\langle H \rangle$.  An example is shown in Table~\ref{tab:mixed}
which shows VMC, mixed, and extrapolated energies for
$^6$Li computed with the AV18+IL2 Hamiltonian.  The sum
of the extrapolated kinetic and potential energy values is
4.3 MeV different from the total energy.  This means that
the extrapolated values of the pieces have errors whose
absolute sum is at least this big.

Instead of the linear extrapolation of Eq.~(\ref{eq:pc_gfmc}), one
can also use a ratio extrapolation:
\begin{eqnarray}
\langle O (\tau)\rangle
\approx  \frac{\langle O (\tau)\rangle_{\rm Mixed}^2}
     {\langle O \rangle_T} ~,
\label{eq:ratio-extrap}
\end{eqnarray}
which is the same as Eq.~(\ref{eq:pc_gfmc}) to lowest order in
$[\langle O (\tau)\rangle_{\rm Mixed} - \langle O \rangle_T]$,
but obviously has different quadratic errors.  This method has
the feature that if both $\langle O (\tau)\rangle_{\rm Mixed}$
and $\langle O \rangle_T$ have the same sign, then the
extrapolated $\langle O (\tau)\rangle$ will also have that sign.
This is an advantage for quantities such as densities
(see Sec.~\ref{sec:dens}) which must be positive; if the
GFMC is reducing the density at large $r$ where it is 
exponentially falling, linear extrapolation can result in
negative values.  Of course such large extrapolations by
either method are uncertain.

\begin{table}[bt]
\centering
\begin{minipage}[t]{3.00in}
\begin{tabular}{lrrcr}
\hline                                                        
 &                $\langle O \rangle_T$ & $\langle O \rangle_{\mbox{Mix}}$ &
 $\langle O \rangle_{\mbox{Mix}}-\langle O \rangle_T$ & $\langle O \rangle~~$ \\
\hline
$K$       &    146.60 &   153.49 &   ~~6.89 &   160.39 \\
$v_{\mbox{\small nuc}}$  &
             --171.64 & --180.49 &   --8.85 & --189.34 \\
$v_{C}$   &      1.53 &     1.54 &   ~~0.01 &     1.55 \\
$V_{ijk}$ &    --4.10 &   --6.43 &   --2.34 &   --8.77 \\
\hline
Sum       &   --27.61 &  --31.90 &   --4.28 &  --36.17 \\
\hline
$H$       &   --27.61 &  --31.90 &     ---  &  --31.90 \\
\hline
\end{tabular}
\caption{Contributions to  $\langle H \rangle$ for $^6$Li  (MeV)}
\label{tab:mixed}
\end{minipage}
\end{table}

\subsection{A Simplified GFMC Calculation}
The basic steps in a GFMC calculation are
\begin{itemize}
\item Start with collection of $\Psi(\tau\!=\!0,{\bf R}_j) = \Psi_T({\bf R}_j)$
from a VMC calculation
\item Loop over time steps $\tau_n = n \triangle \tau$
\begin{itemize}
\item Loop over configurations $j$
\begin{itemize}
\item Make a random step to ${\bf R}_j^\prime =  {\bf R}_j + \triangle {\bf R}_j $
by sampling $G_{0}({\bf R}^{\prime},{\bf R})$
\item Sample several directions based on simplified $\Psi_T$ and potential
\item Compute $\Psi(\tau_{n+1},{\bf R}_j^{\prime}) =  
G({\bf R}_j^{\prime},{\bf R}_j) \Psi(\tau_{n},{\bf R}_j) $
\item Possibly mark as killed due to the constraint  
$\Psi^\dagger(\tau_{n+1},{\bf R}_j^\prime) \cdot \Psi_T({\bf R}_j^\prime)$
\item Use importance sampling to kill or replicate the configuration 
$\Psi(\tau_{n+1},{\bf R}_j^{\prime})$
\end{itemize}
\item Every 20--40 time steps
\begin{itemize}
\item Compute the local energy
$\Psi^\dagger(\tau_{n},{\bf R}_j) H \Psi_T({\bf R}_j)
/\Psi^\dagger(\tau_{n},{\bf R}_j) \Psi_T({\bf R}_j)$
\item Check that total number of configurations is staying reasonably constant
\end{itemize}
\end{itemize}
\end{itemize}

GFMC calculations are quite computer intensive.  For example, a typical
$^8$Li calculation requires 300 processor hours running at a delivered
(not theoretical-peak) speed of one GigaFLOPS.  As shown in
Table~\ref{tab:psi_time}, a $^{10}$B calculation will need about 10
times this and $^{12}$C 250 times it.  Clearly such calculations are
practical only on highly parallel computers.  Our GFMC program uses a
master-slave structure in which each slave gets a number of configurations to
propagate as outlined in the preceding paragraph.  The computed
energy results are sent back to master for averaging as they are generated.

Because configurations are multiplied or killed during propagation, 
the work load fluctuates.  It is important to periodically 
rebalance the work load -- otherwise slaves will wind up with nothing
to do while the last slave with the biggest work load finishes its
calculations.  To do this, the master periodically collects load statistics 
and then tells slaves to redistribute some of their configurations.
The slaves have work (energy calculations left from previous time steps) 
set aside to do during this synchronization.
This method results in 
parallelization efficiencies of typically 95\% on up to 2000 processors.
However, the next generation of large computers will have 10,000 to 100,000
processors, which is more than the number of configurations to
be propagated for a large nucleus like $^{12}$C.  Thus the program
has to be made parallel at a finer level; this is being worked on.

\subsection{Examples of GFMC propagation}

\begin{figure}[bt] 
\includegraphics[height=3.0in]{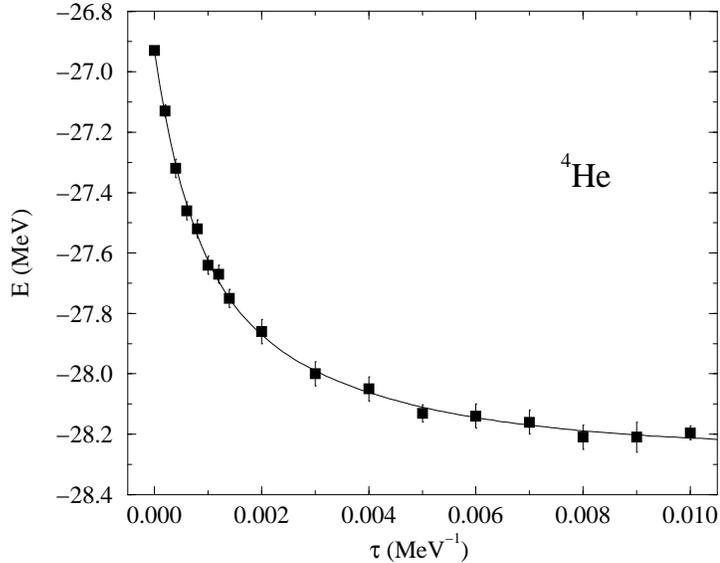}
\caption{GFMC propagation for $^4$He.  The $E(\tau)$ is shown
as a function of imaginary time, $\tau$.}
\label{fig:4hegfmc}
\end{figure}

Figure~\ref{fig:4hegfmc} shows the $E(\tau)$ as a function of
the imaginary time, $\tau$, for the beginning of GFMC propagation for $^4He$.
The propagation starts from the VMC value of --26.92~MeV,
and initially decreases rapidly with increasing $\tau$;
essentially converged values are achieved by $\tau=0.01$~MeV$^{-1}$.
The propagation is continued another factor of 10 to 
$\tau=0.1$~MeV$^{-1}$ and the results averaged over the last
half of the propagation to get the converged result of --28.300(15)~MeV.
The curve is a fit of the form
\begin{equation}
E(\tau) = E_0 + \frac{ \sum_{i} \alpha_i^2 E_i^\star exp(-E_i^\star \tau) }
{1 + \sum_{i} \alpha_i^2 exp(-E_i^\star \tau) } \ ,
\end{equation}
to the computed $E(\tau)$ using three terms.  The $E_1^\star$ was
fixed at the first 0$^+$ excitation energy of 20.2~MeV of $^4$He,
and the other two $E_i^\star$ and the three $\alpha_i$ were varied
in the fit.  The fitted $E_i^\star$ turn out to be very large,
340 and 1480~MeV, with small $\alpha_i$, 0.0018 and 0.00046, respectively.
Thus the errors in the VMC $\Psi_T$ correspond to small amounts
of extremely high excitation energy; GFMC is particularly efficient
at filtering out such errors.

\begin{figure}[bt] 
\includegraphics[height=5.0in,angle=270]{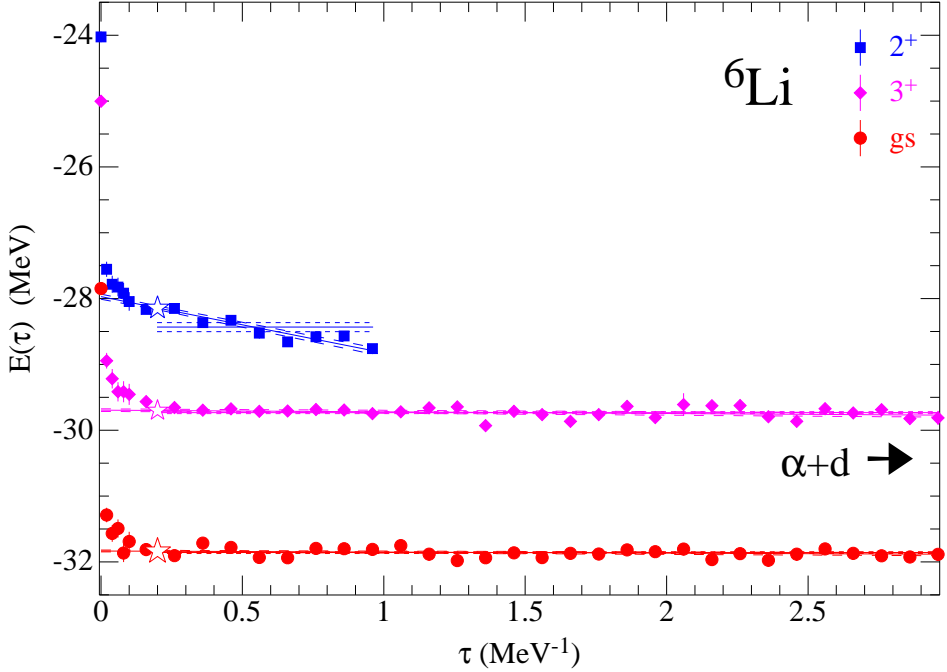}
\caption{GFMC propagation for three states of $^6$Li.}
\label{fig:6liegfmc}
\end{figure}

Figure~\ref{fig:6liegfmc} shows GFMC propagation, using the AV18+IL2
Hamiltonian, of the ground, first 3$^+$, and 2$^+$ states of $^6$Li.
The propagation for the ground state (which is particle stable with this
$H$) and the 3$^+$ (which is only slightly above the d+$\alpha$
threshold and experimentally has a narrow width) is stable after $\tau$
= 0.2 MeV$^{-1}$.  However the 2$^+$ (a broad resonance) never becomes
stable; the $E(\tau)$ are decaying to the threshold energy of separated
$\alpha$ and d clusters.  Because the 3$^+$ state $E(\tau)$ stops
decreasing around $\tau$=0.2, the $E(\tau$=0.2), shown by the star, is
best GFMC estimate we can currently make of the resonance energy.
However it is now possible to make GFMC calculations using
scattering-wave boundary conditions (see Sec.~\ref{sec:scat}) and
this method will be applied to states such as $^6$Li(2$^+$).

\begin{figure}[bt] 
\centering
\includegraphics[height=5.0in,angle=270]{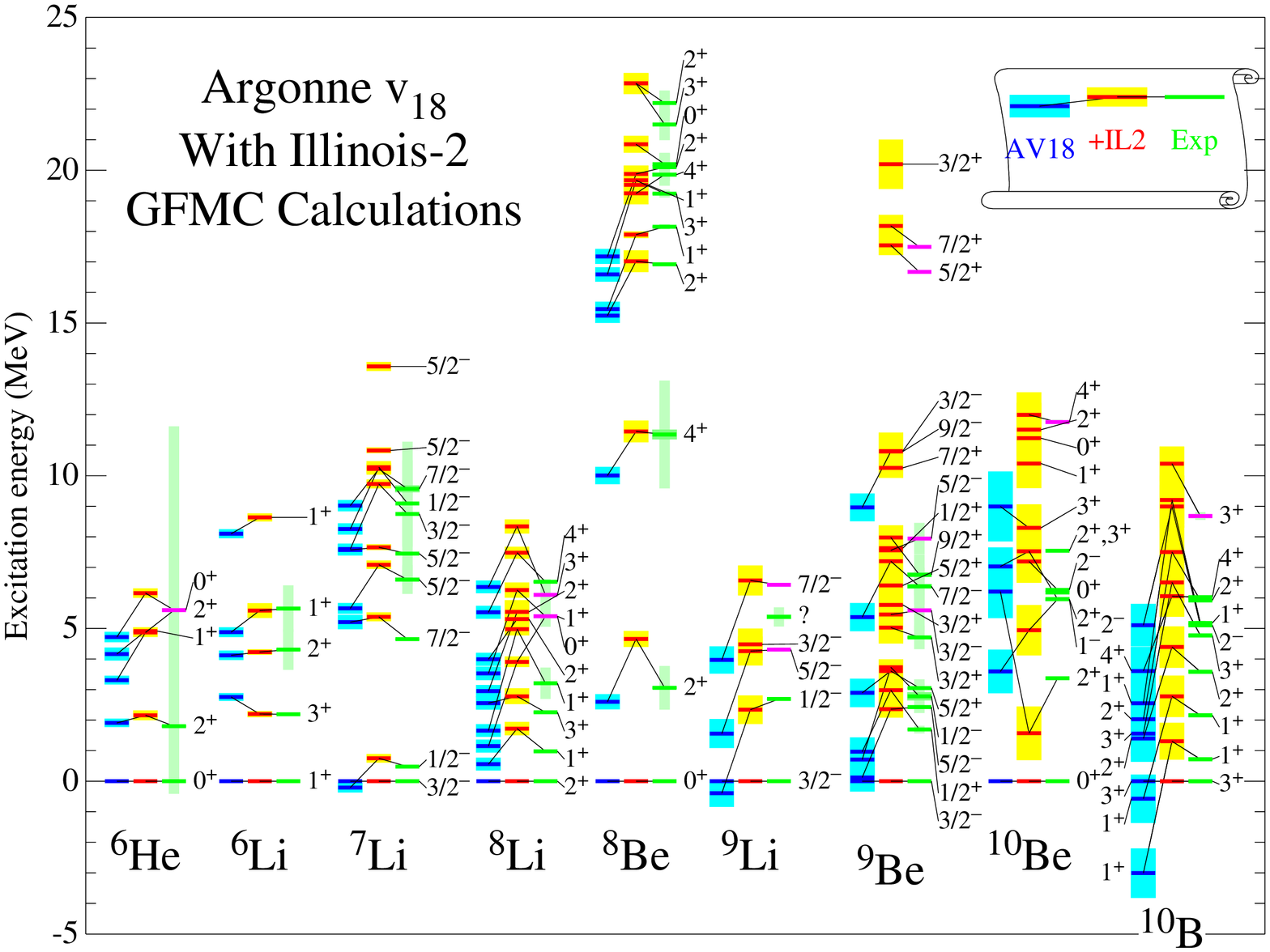}
\caption{GFMC computations of excitation energies for the AV18 and AV18+IL2 Hamiltonians 
compared with experiment.  The shading on the experimental energies shows
the widths of resonances.}
\label{fig:estar-av18-il2-exp}
\end{figure}

\section{Results for energies of nuclear states}
\label{sec:energy}

Figure \ref{fig:av18-il2-exp} compares the GFMC energies of various nuclear
states with experiment and Fig.~\ref{fig:estar-av18-il2-exp} does the
same for excitation energies.  In both cases the left set of bars for
each isotope shows results using just the AV18 \NN potential while
the middle set of bars is for the full AV18+IL2 Hamiltonian.  As has
already been observed, AV18 alone significantly underbinds all nuclei
except the deuteron; including IL2 results in fairly good agreement
with the experimental values.  The excitation spectra in Fig.~\ref{fig:estar-av18-il2-exp}
show that IL2 also fixes other problems that arise when just a \NN
potential is used.  For example, spin-orbit splittings are usually
too small without the \NNN potential (note the 
$\case{1}{2}^--\case{3}{2}^-$ and $\case{5}{2}^--\case{7}{2}^-$ splittings
in $^7$Li and the $\case{1}{2}^--\case{3}{2}^-$ splitting in $^9$Li).
As is discussed in the next subsection, even the ordering of
states can be changed by the \NNN potential.

The discussion of Sec.~\ref{sec:gfmc} implies that GFMC can be
used only for the lowest state of each set of quantum numbers
but Fig.~\ref{fig:av18-il2-exp} shows several states with the
same $J^\pi$.  The ability of GFMC to provide such results
was demonstrated in Ref.~\cite{PWC04}.

\subsection{Ordering of States in $^{10}$Be and $^{10}$B}

\begin{figure}[bt] 
\includegraphics[height=4.00in,angle=270]{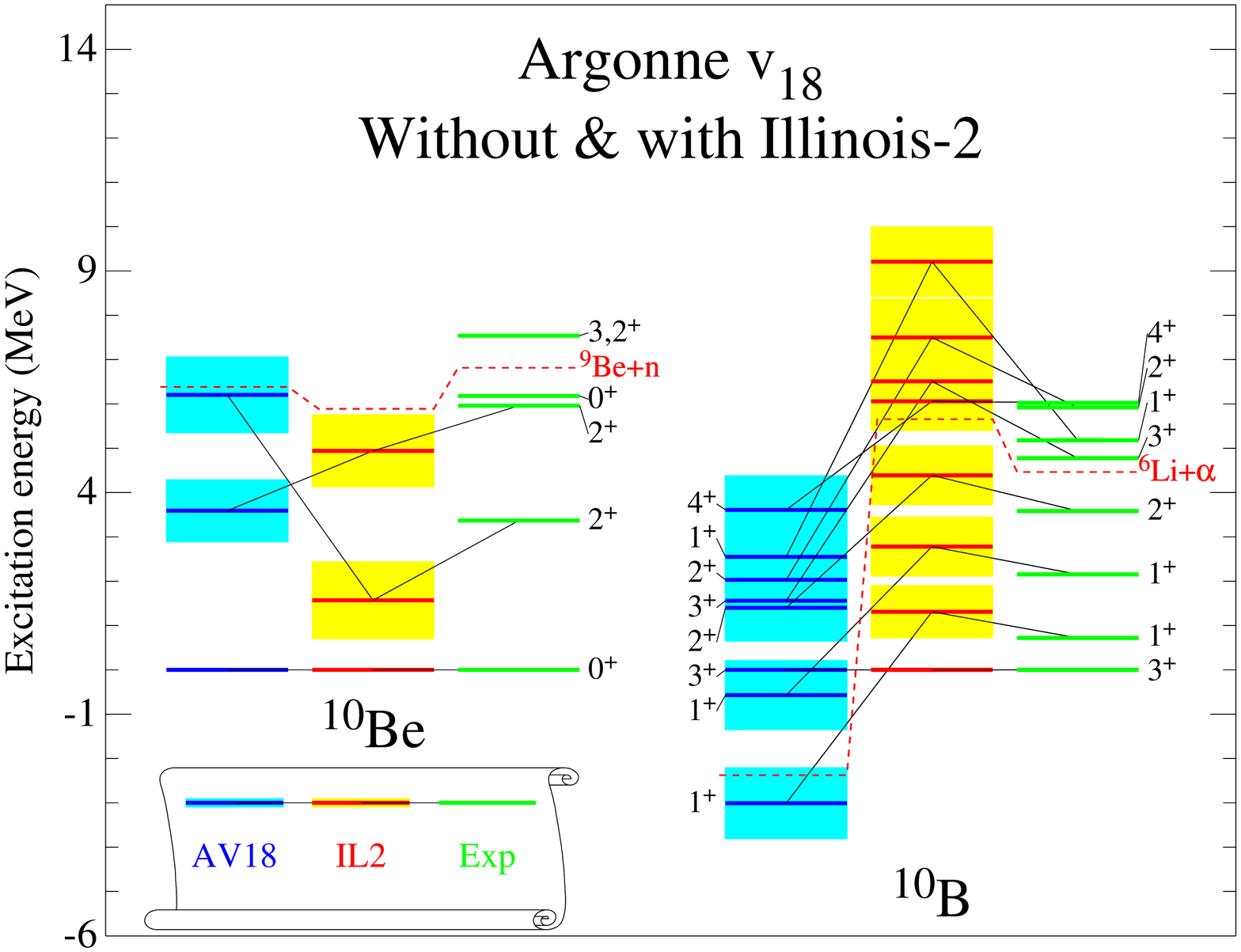}
\caption{Excitation energies of $^{10}$Be and $^{10}$B}
\label{fig:estar-10be-10b}
\end{figure}

Figure~\ref{fig:estar-10be-10b} shows the beginning of the 
computed and experimental excitation
spectra of $^{10}$Be and $^{10}$B.  We see that 
$N\!N$ potentials with no $N\!N\!N$ predict a 1$^+$ ground state for $^{10}$B 
while the Illinois-2 $N\!N\!N$ potential fixes this and gives 
the correct 3$^+$ ground state.
No-core shell model calculations show that other $N\!N$ potentials
without \NNN potentials also give a 1$^+$ ground state for $^{10}$B~\cite{NCSM-10B},
so this is not a failure of just AV18.  This incorrect ground-state
prediction is another manifestation of too-small spin-orbit splitting
using just \NN potentials; in 1956 D. Kurath showed that the relative
positions of the 3$^+$ and 1$^+$ levels depends on the amount
of spin-orbit strength in a shell-model calculation~\cite{K56}.

The first two excited states in $^{10}$Be are both 2$^+$ and
Fig.~\ref{fig:estar-10be-10b} shows that including the IL2 \NNN
potential reverses their order.  
VMC and GFMC calculations predict large positive and negative quadrupole moments ($Q$)
for these states; with no \NNN potential, the GFMC energy of the $Q>0$ state
is the lower, while adding the IL2 \NNN potential reverses this.
VMC also predicts a large B(E2) to the g.s. for only the state with $Q>0$.
An ATLAS experiment for the B(E2) and quadrupole moments of these states will be made
to determine if the reversal of order given by IL2 is correct.


\subsection{Charge Dependence and Isospin Mixing}

The differences of the energies of states in the same isomultiplet are
a probe of isospin-breaking components of the Hamiltonian.  The largest
such component is the Coulomb potential between protons, but it has
been known since 1969 that this does not fully account for the
measured differences~\cite{NS69}; the discrepancy is referred to as
the Nolen-Schiffer anomaly.  It is convenient to parametrize the
energies of the isomultiplets by the coefficients $a^{(n)}_{A,T}$,
\begin{eqnarray}
   E_{A,T}(T_z) = \sum_{n\leq 2T} a^{(n)}_{A,T} Q_n(T,T_z) \ ,
\end{eqnarray}
where $Q_0=1$ is the isoscalar component, $Q_1=T_z$ the isovector, and
$Q_2=\case{1}{2}(3T_z^2-T^2)$ the isotensor.  Term 18 of AV18 contributes
to the isovector component and terms 15 to 17 to the isotensor component;
both terms receive contributions from the various electromagnetic
terms in AV18.  Table~\ref{table:ns} shows computed and experimental
values of the coefficients for several isomultiplets; 
the $v^{\rm CSB+CD}$ column gives contribution of the nuclear (strong-interaction) terms
in AV18, while the $K^{\rm CSB}$ column shows that resulting from the difference
of the proton and neutron masses.
In general the
non-Coulomb electromagnetic and strong charge symmetry breaking and charge dependent
terms result in good agreement with the experimental values and thus resolve
the Nolen-Schiffer anomaly.

\begin{table}
\begin{tabular}{cccrcccrr}
\hline
& $T$ & $n$ & $v^{\rm Coul}$  & $v^{\rm other EM}$ & $v^{\rm CSB+CD}$ & $K^{\rm CSB}$  
                                                                      & Total   & Expt. \\
\hline

$^3$H--$^3$He                       &$\case{1}{2}$ & 1 &   649 &  29 &  64 & 14 &  757 &  ~764 \\
\hline                                                        
$^7$Li--$^7$Be                      &$\case{1}{2}$ & 1 &  1458 &  40 &  83 & 23 & 1605 &  1644 \\
\hline                                                        
$^7$He,$^7$Li$^*$,$^7$Be$^*$,$^7$B  &$\case{3}{2}$ & 1 &  1286 &  14 &  49 & 17 & 1366 &  1373 \\
                                    &$\case{3}{2}$ & 2 &   132 &  ~7 &  34 &    &  174 &  ~175 \\
\hline                                                        
 $^8$Li,$^8$Be$^*$,$^8$B            &    1         & 1 &  1692 &  24 &  78 & 24 & 1818 &  1770 \\
                                    &    1         & 2 &   140 &  ~5 &$-$5~&    &  140 & ~~145 \\
\hline                                                        
$^8$He,$^8$Li$^*$,$^8$Be$^*$,$^8$B$^*$,$^8$C  
                                    &    2         & 1 &  1719 &  13 &  83 & 26 & 1840 &  1659 \\
                                    &    2         & 2 &   153 &  ~7 &  42 &    &  203 & ~~153 \\
\hline
\end{tabular}
\caption{Computed (AV18+IL2) and experimental Nolen-Schiffer energies for several isomultiplets}
\label{table:ns}
\end{table}

\begin{figure}[bt] 
\centering
\includegraphics[height=3.10in]{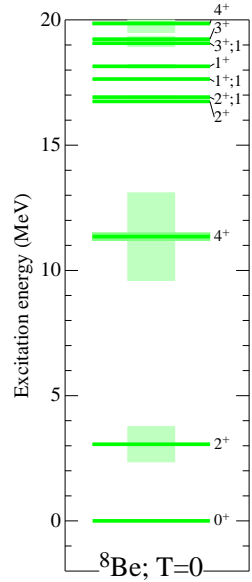}
\caption{Experimental spectrum of $^8Be$}
\label{fig:8be-spec}
\end{figure}

Figure \ref{fig:8be-spec} shows the beginning of the experimental excitation
spectrum of $^8$Be.  From 16 to 19 MeV excitation there are pairs of 2$^+$,
1$^+$, and 3$^+$ levels with isospin 0 and 1.  The two 2$^+$ levels are 
very close and hence strongly isospin mixed; the 3$^+$ levels also have
significant mixing.  The mixing has been experimentally known from the decay
properties of the states since 1966~\cite{B66}.  However, as with
the Nolen Schiffer anomaly, calculations using just Coulomb mixing
underestimate the amount of mixing.  Table~\ref{table:isomix} shows
GFMC calculations using AV18+IL2 of the mixing matrix elements for
the three pairs~\cite{8be-mix}.  The contribution of the nuclear CSB term is relatively
more important here than for the Nolen Schiffer anomaly.  The agreement
of the predictions with the data is not as satisfactory as for the
Nolen Schiffer anomaly.  The final line of the table shows the
mixing matrix element between the first 2$^+$ (at 3 MeV) state and the
isospin-1 17-MeV state.  The small value of this matrix element,
and the large energy denominator, shows that the first 2$^+$ state
has very little $T=1$ contamination; this is important to
the possibility of using $^8$Li($\beta^-$)$^8$Be(1$^{st}$ 2$^+$) decay
as a test of $V-A$.

\begin{table}[bt]
\begin{tabular}{cccccc}
\hline                                                        
$ J^P$          & \multicolumn{4}{c}{GFMC}                  &   Expt \\
               &  Coulomb &  Strong CSB &  Other &   Total &        \\
\hline
 2$^{nd}$ 2$^+$ &   78     &    21~~     &   16   &    115  &    144 \\
          1$^+$ &   80     &    18~~     &   ~4   &    102  &    120 \\
          3$^+$ &   61     &    15~~     &   14   &    ~90  &    ~63 \\
\hline
 1$^{st}$ 2$^+$ &   ~4     &    ~0.4     &   ~1   &    ~~6  &    --  \\
\hline
\end{tabular}
\caption{Isospin mixing matrix elements for $^8$Be in keV}
\label{table:isomix}
\end{table}


\subsection{Can Modern Nuclear Hamiltonians Tolerate a Bound Tetraneutron?}

\begin{figure}[bt] 
\includegraphics[height=5.0in,angle=270]{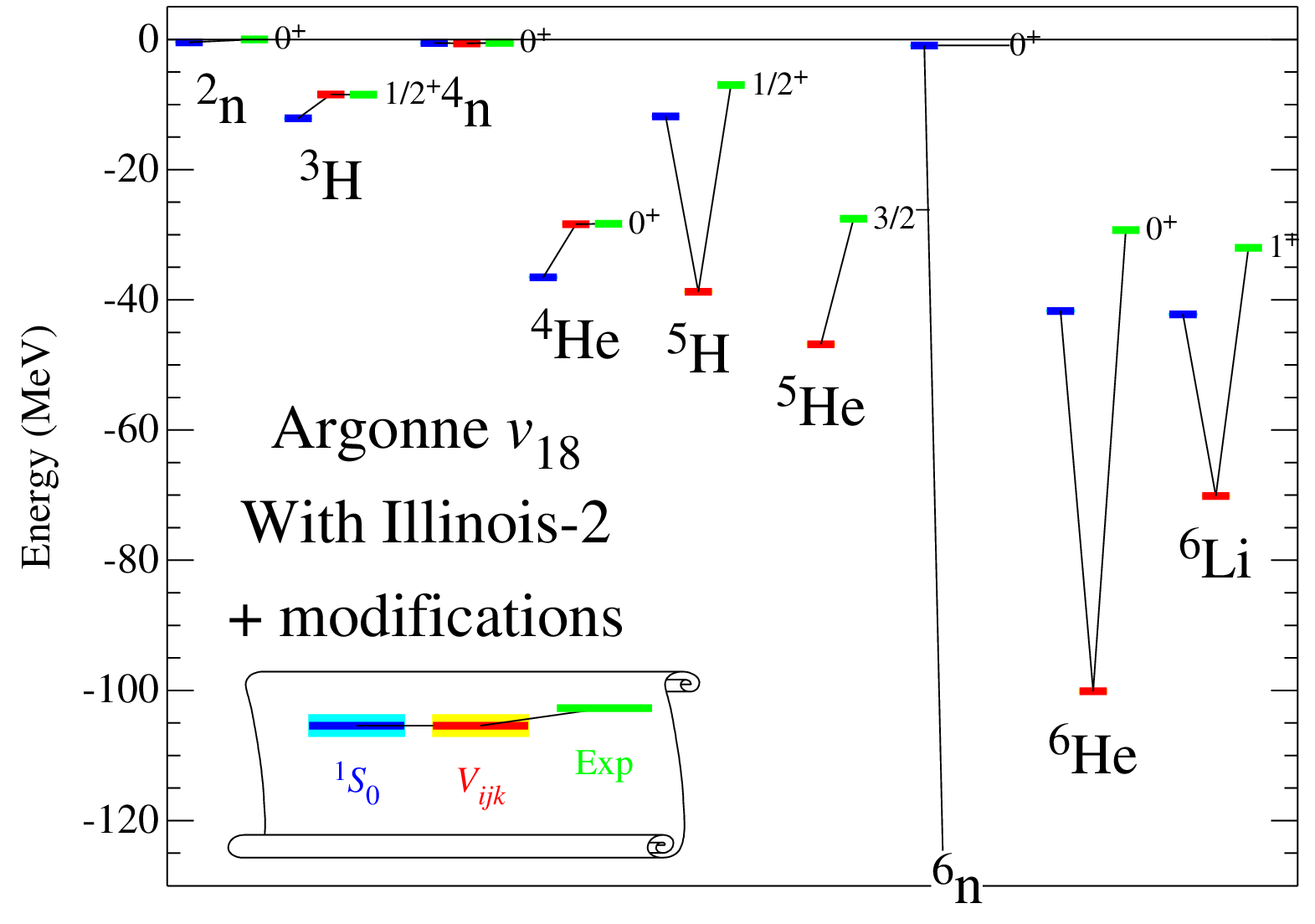}
\caption{Nuclear energies with Hamiltonians modified to bind a tetraneutron}
\label{fig:4n}
\end{figure}

In 2002 a claimed observation of a bound tetraneutron was
published~\cite{4n-exp,4n-exp-rev}.  The experiment did not produce a
definite binding energy, just the statement that four neutrons are weakly
bound.  It is well known that the dineutron is not bound, but rather has
low-energy pole on the second sheet (pseudo bound state); AV18 along
with other realistic potentials has this feature.  A set of GFMC
calculations were made to see if the AV18+IL2 Hamiltonian, or acceptable
modifications of it, could reproduce the tetraneutron claim~\cite{P03}.
It was clearly established that the unmodified AV18+IL2 does not bind
$^4$n; at most there is some weak resonance at $E \sim +2.$ MeV.

Minimal modifications to AV18+IL2 to give E($^4$n) $\sim -$0.5 MeV were
then made and the effects of such modifications on the energies of
other, well established, nuclei were computed.  Figure~\ref{fig:4n}
shows some of these.  In the first case (the left-hand bars), the
intermediate-range part of the $^1\!S_0$ partial-wave potential in of AV18
was increased enough to bind four neutrons.  This increase results in
the dineutron also being bound, in fact the $^4$n is still unbound
against breakup into two dineutrons in this model.  As the figure shows,
other existing nuclei ($^3$H, $^4$He, etc.) all become significantly
over bound with this increased $^1\!S_0$ potential.  Also $^6$n and $^8$n
(not shown) become bound.  An attempt to bind the $^4$n by changing
the $^3P_J$ part of AV18 was also made, but this requires a huge
change which very strongly overbinds other nuclei.

A second attempt was to add an attractive $V_{ijk}(T=\case{3}{2})$ to $H$.
This has two advantages:  1) it has no effect on $N\!N$ scattering
and does not make a bound $^2$n;  2) because the modification is made only
in isospin-$\case{3}{2}$ triples, it has no effect on $^3$H, $^3$He, or $^4$He;
these have only isospin-$\case{1}{2}$ triples.  However, as can be
seen in the figure, as soon as this potential can act (i.e. in nuclei
with $T=\case{3}{2}$ triples), it produces huge overbinding.  The most
dramatic effect is in pure neutron systems: $^6$n is bound by 220 MeV and
$^8$n by 650 MeV.  These are the most stable 6- and 8-nucleon systems
with this Hamiltonian, so all other 6- and 8-body nuclei would beta
decay to them!

Finally an attractive four-nucleon, $T=2$, potential was added (not shown).  This
does not effect $^6$Li but does very strongly overbind $^6$He, larger
nuclei, and pure neutron systems with more than four neutrons.  The
conclusion of the study is that a bound $^4$n is incompatible with
our understanding of nuclear forces.  In the meantime the experiment
has not been successfully reproduced.

\begin{figure}[bt] 
\begin{minipage}[t]{2.60in}
\includegraphics[height=2.70in,angle=270]{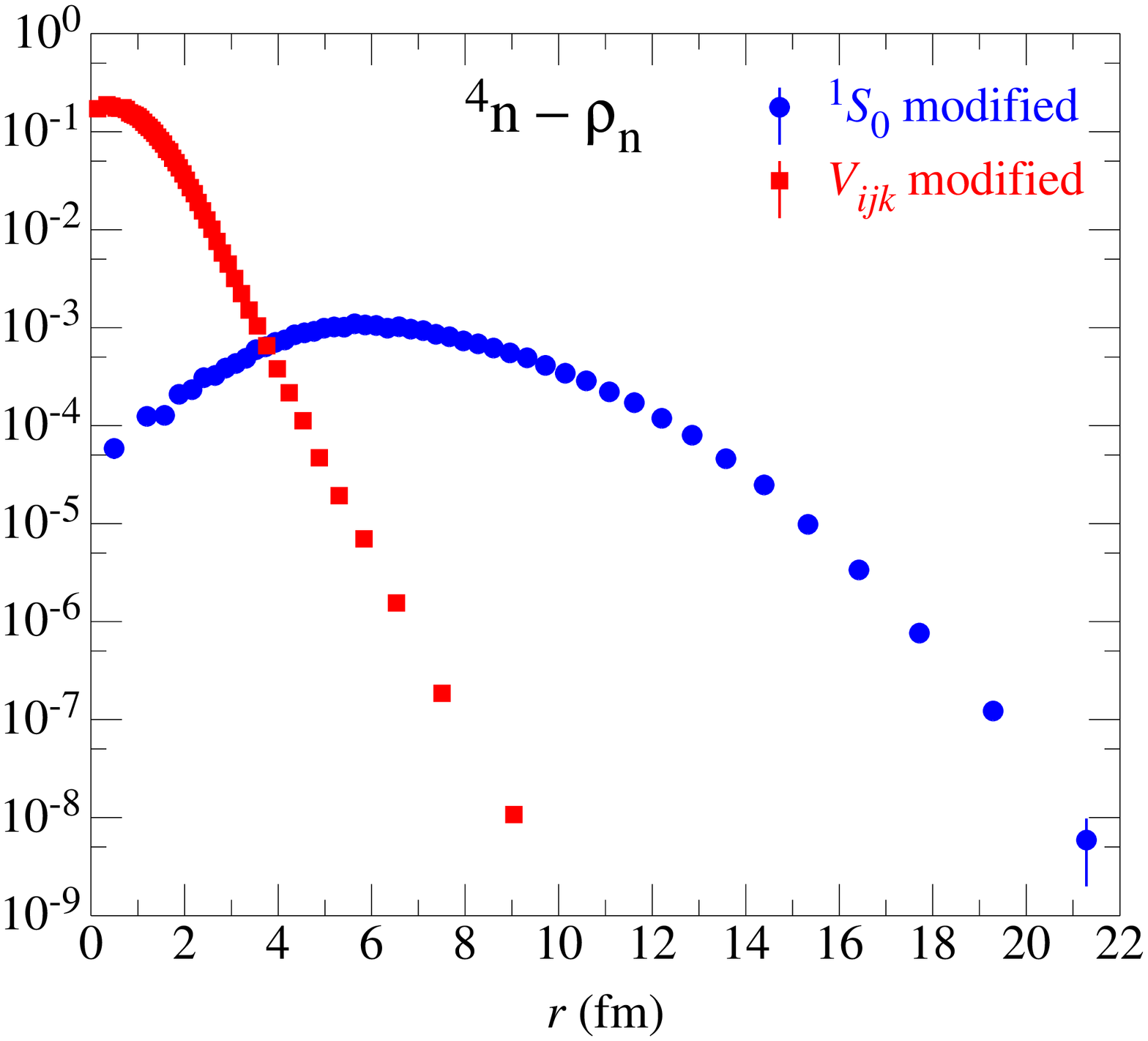}
\end{minipage}
\begin{minipage}[t]{2.50in}
\includegraphics[height=2.70in,angle=270]{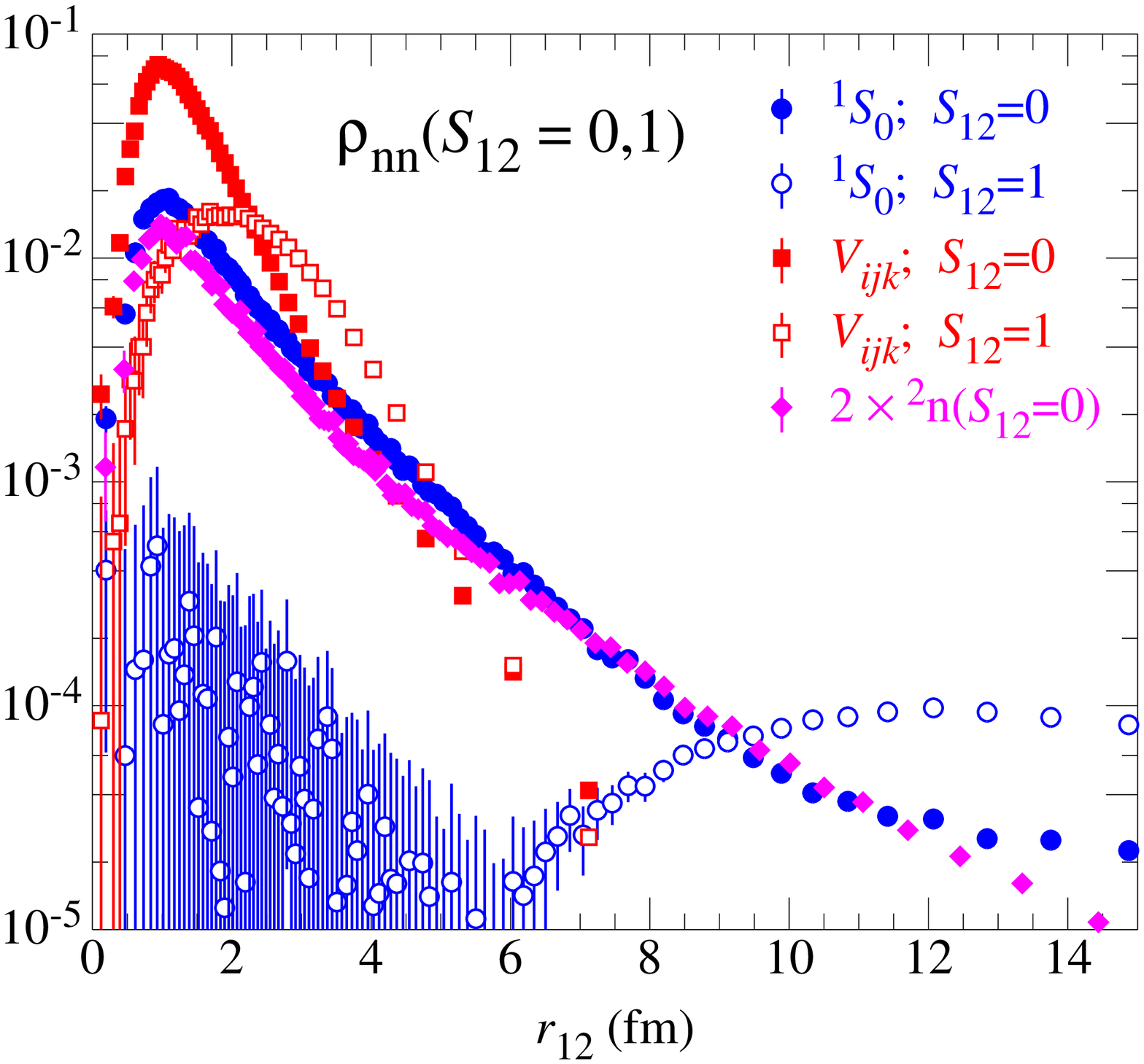}
\end{minipage}
\caption{One-body (left) and two-body (right) densities of tetraneutrons.
Circles are for the $^1\!S_0~N\!N$ and squares for the
$T=\case{3}{2}~N\!N\!N$ modifications; the diamonds are for $^2$n.}
\label{fig:4n-rho}
\end{figure}

The $^1\!S_0~N\!N$ and $T=\case{3}{2}~N\!N\!N$
modifications have very different effects on $A>4$ binding energies,
even though both have been adjusted to bind $^4$n by only 0.5 MeV;
the \NNN modification results in much more severe overbinding.
Figure~\ref{fig:4n-rho} helps to explain this.  The left panel
shows the one-body densities of the two bound $^4$n systems; 
the $^1\!S_0~N\!N$ modification results in a very diffuse $^4$n
with a rms radius of 8.9 fm.  The right panel shows the pair
density which is proportional to finding two neutrons a given
distance apart.  For the $^1\!S_0~N\!N$ modification, the density
for a $S$=0 pair
is peaked around 1~fm and is very similar to the pair distribution
of the isolated $^2$n which is bound by this potential.  
The $S$=1 pair distribution is peaked at 12~fm; it arises from
neutrons in different dineutron clusters.  Thus
this $^4$n looks like two widely separated dineutrons.  The 
binding comes from the small tails where neutrons from each
dineutron get close enough to interact.  The change of 
$v_{N\!N}(^1\!S_0)$ needed to achieve this is not that big
and hence bigger nuclei overbound only somewhat.

On the other hand, a $V_{N\!N\!N}$ requires all three neutrons
in a triple to be close together to be effective.  Thus the
$T=\case{3}{2}~N\!N\!N$ modification must bring the two
dineutrons close together.  This results in the much more
compact one-body density, with a rms radius of only 1.9 fm,
shown in the left panel and a two-body density that is much
sharper than the isolated $^2$n.  
The $V_{N\!N\!N}(T\!=\!\case{3}{2})$ must be large to achieve
this high density and it thus has a large effect in all bigger
nuclei.


\section{GFMC for Scattering States}
\label{sec:scat}

The GFMC calculations presented so far have treated the nucleus as a
particle-stable system; that is the starting wave function, $\Psi_T$, is
exponentially decaying as any nucleon is removed to a large distance from
the center of mass.  However many of the states of interest are particle
unstable; they are above the threshold for emission of a single nucleon
or, as is often the case for light nuclei, the threshold for breakup
into subclusters.  This approximation appears to be adequate for narrow
resonances which have only a small scattering-wave component.  However broad
resonances should really be computed with proper scattering-wave
boundary conditions; as shown in Fig.~\ref{fig:6liegfmc} we do not
achieve a converged energy for a broad state with the bound-state GFMC.
In addition to being the correct approach, scattering solutions allow
one to compute the phase shift as a function of energy and thus obtain
the width of the resonance.  Finally, one might also be interested in the
phase shifts for partial waves that have no resonance.

\begin{figure}[bt] 
\includegraphics[height=4.5in,angle=270]{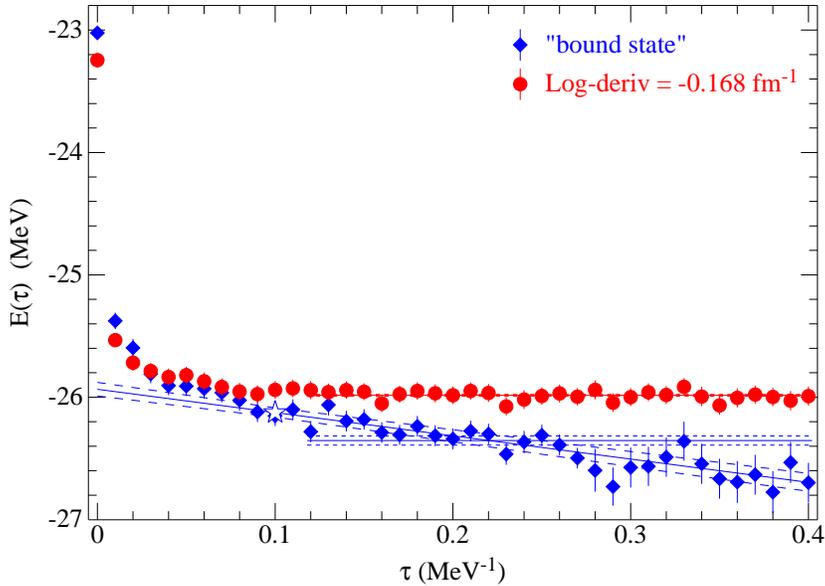}
\caption{GFMC propagation for $^5$He($\case{1}{2}^-$) using
bound- and scattering-state boundary conditions (diamonds and circles,
respectively).}
\label{fig:scat-tau}
\end{figure}

Preliminary GFMC calculations of neutron-alpha scattering were made 
in 1991~\cite{C91}, but detailed, high statistics results have been
obtained only recently~\cite{NPWCH07}.  Instead of using exponentially
decaying wave functions, we construct $\Psi_T$ to have a specified 
logarithmic derivative, $\gamma$, at some large 
boundary radius ($R \ge 7$ fm).  Here $R$ is the maximum distance
that any nucleon is allowed to get from the other $A-1$ nucleons.
The GFMC propagation uses a method of images to preserve 
$\gamma$ at $R$, and thus finds $E(R,\gamma)$; the eigenenergy that
corresponds to the boundary condition.
The phase shift, $\delta(E)$, can then be computed from $R$, $\gamma$, 
and $E(R,\gamma)$.  This procedure is repeated for a number of $\gamma$ 
and $\delta(E)$ is mapped out parametrically.

Figure \ref{fig:scat-tau} shows an example of this for the broad
$\case{1}{2}^-$ resonance in $^5$He.
The bound-state boundary condition does not give a stable energy,
but is decaying to the $n$+$^4$He threshold energy.  
The scattering boundary condition produces a stable energy; the
value of $\gamma$ used in this example results in an energy slightly above the
resonance energy.

\begin{figure}[bt] 
\begin{minipage}[t]{2.50in}
\includegraphics[width=2.50in]{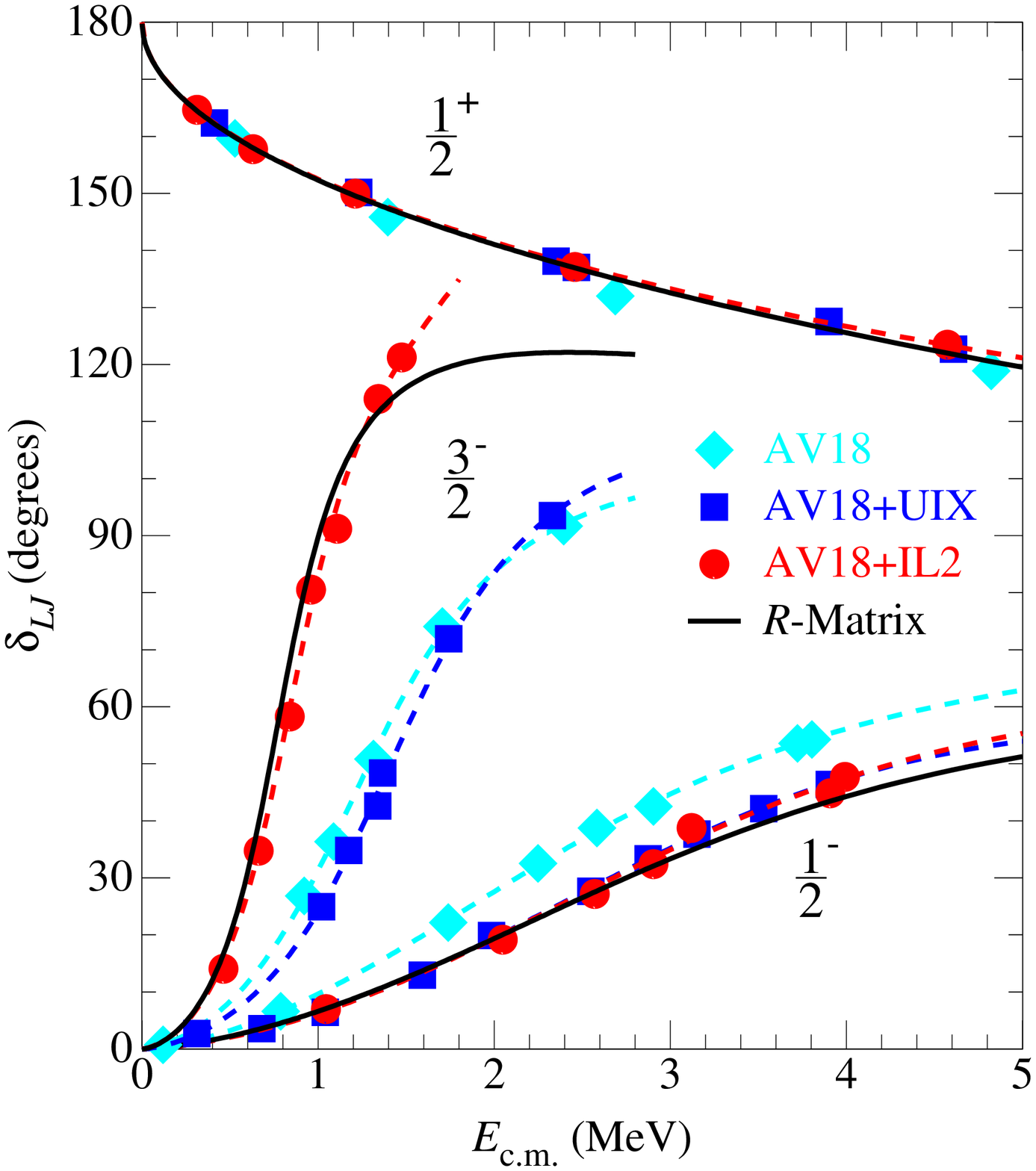}
\end{minipage}
\begin{minipage}[t]{2.50in}
\includegraphics[width=2.60in]{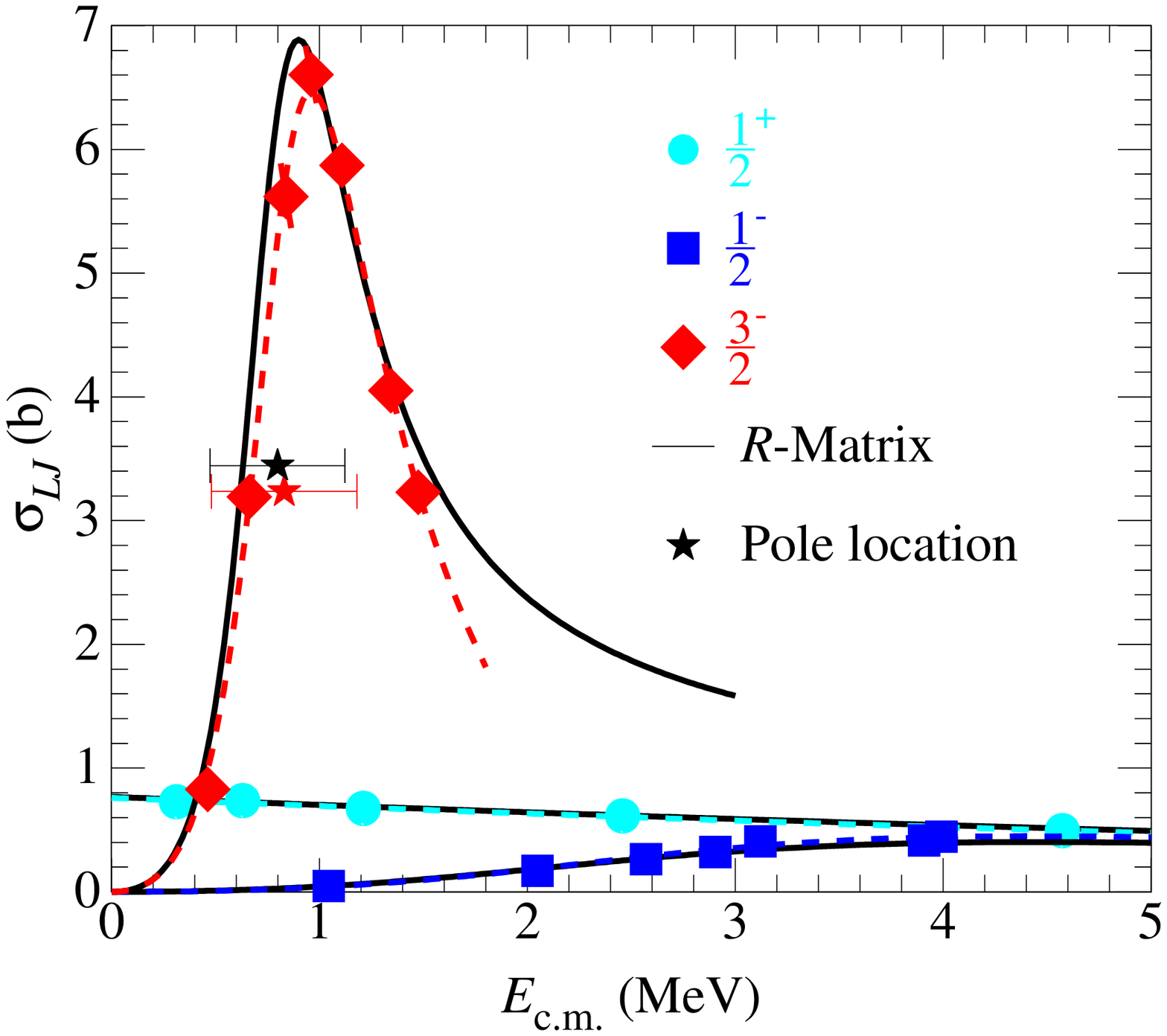}
\end{minipage}
\caption{GFMC calculations of $n+\alpha$ scattering in the three
principal partial waves.  The left panel shows partial-wave phase
shifts for three Hamiltonians.  The right panel shows corresponding
partial-wave cross sections for the AV18+IL2 Hamiltonian.  The experimental
data is represented by the solid curves.}
\label{fig:he5-scat}
\end{figure}

Figure \ref{fig:he5-scat} compares the calculations of $n+\alpha$ scattering
with a $R$-matrix analysis of the data~\cite{hale} (solid curves).  The
left panel shows the partial-wave phase shifts computed for three
different Hamiltonians: AV18 with no $V_{ijk}$, AV18+IL2, and AV18+UIX (UIX
is an older \NNN potential~\cite{PPCW95} that, with AV18, correctly
binds $^3$H and $^4$He, but underbinds $P$-shell nuclei).  All three
Hamiltonians give very similar results for the $\case{1}{2}^+$ partial
wave which has no resonance.  However only the AV18+IL2 correctly
reproduces the two $P$-wave partial waves; AV18 alone misses both
of them and AV18+UIX fits the $\case{1}{2}^-$ partial wave but
has too-small spin-orbit splitting and misses the $\case{3}{2}^-$ one.
The right panel shows the partial-wave cross sections for the
AV18+IL2 Hamiltonian.  The strong resonance in the $\case{3}{2}^-$
channel is very well reproduced showing that both the position
and width of the resonance agree with the data; the much broader
$\case{1}{2}^-$ resonance is also well reproduced.  In addition the
good agreement with the low-energy $\case{1}{2}^+$ cross section
data shows that the scattering length is reproduced.

This first study is very promising; the GFMC method, with its
ability to have correct asymptotic forms, should be applied to
other scattering calculations including a number of broad resonances
[$^{7,9}$He, $^6$Li(2$^+$), $^8$Be(2$^+$,4$^+$), {\it etc.}]
and the initial states of astrophysically interesting capture reactions
[$^4$He(d,$\gamma$)$^6$Li, $^7$Be(p,$\gamma$)$^8$B, {\it etc.}].


\section{Coordinate- and momentum-space densities}
\label{sec:dens}

Up to now we have been concentrating on GFMC calculations
of energies of nuclear states.  However matrix elements of
any operator may be evaluated in the GFMC propagation by using
the extrapolation formulas, Eqs.~(\ref{eq:pc_gfmc}) or
(\ref{eq:ratio-extrap}).  I discuss some recent work on the
charge radii, the corresponding densities, and momentum-space
densities in this section.

\subsection{RMS radii and one-body densities of helium isotopes}

\begin{figure}[bt] 
\begin{minipage}[b]{2.75in}
\includegraphics[height=2.80in,angle=270]{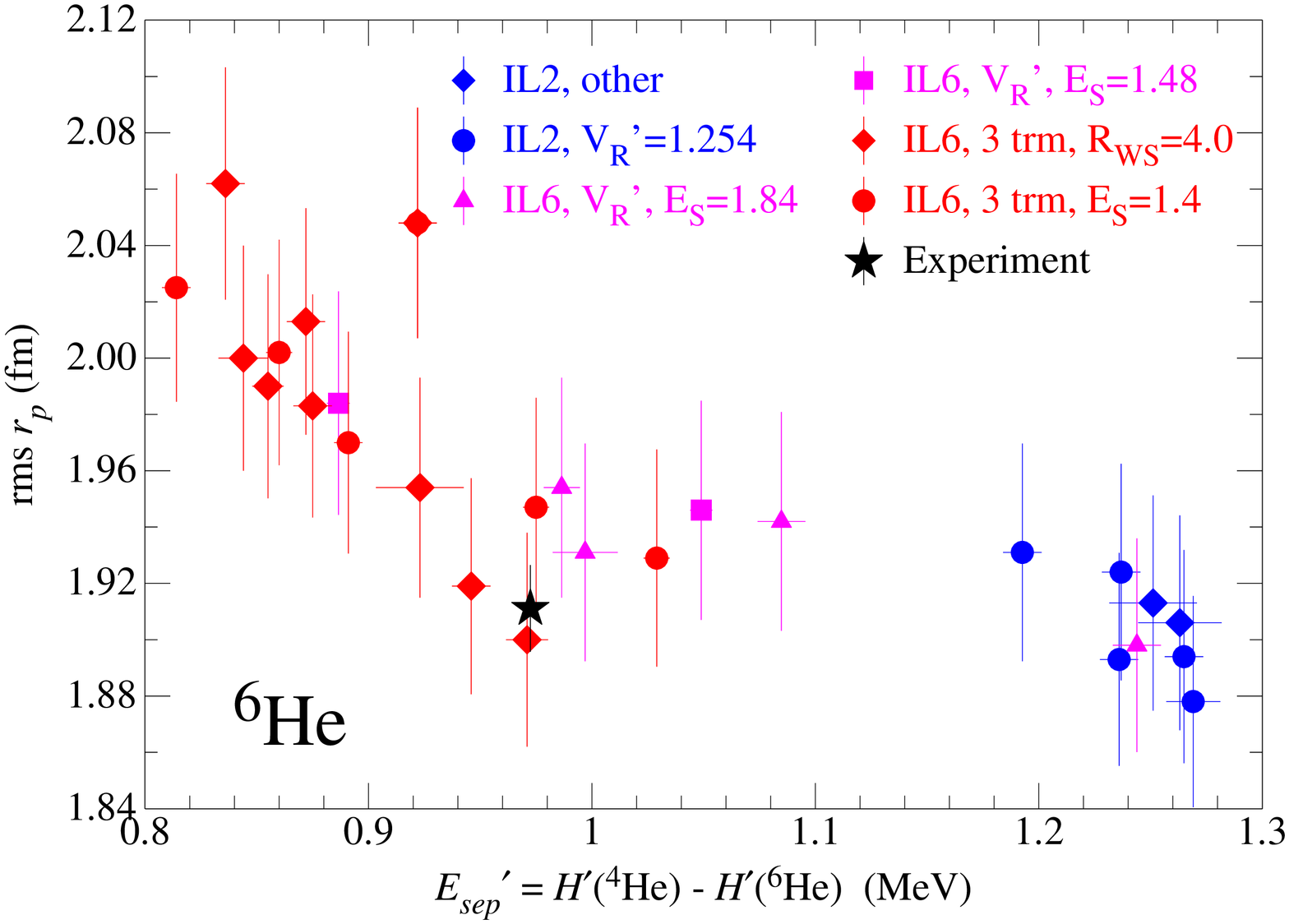}
\end{minipage}
\begin{minipage}[b]{2.50in}
\includegraphics[height=2.60in,angle=270]{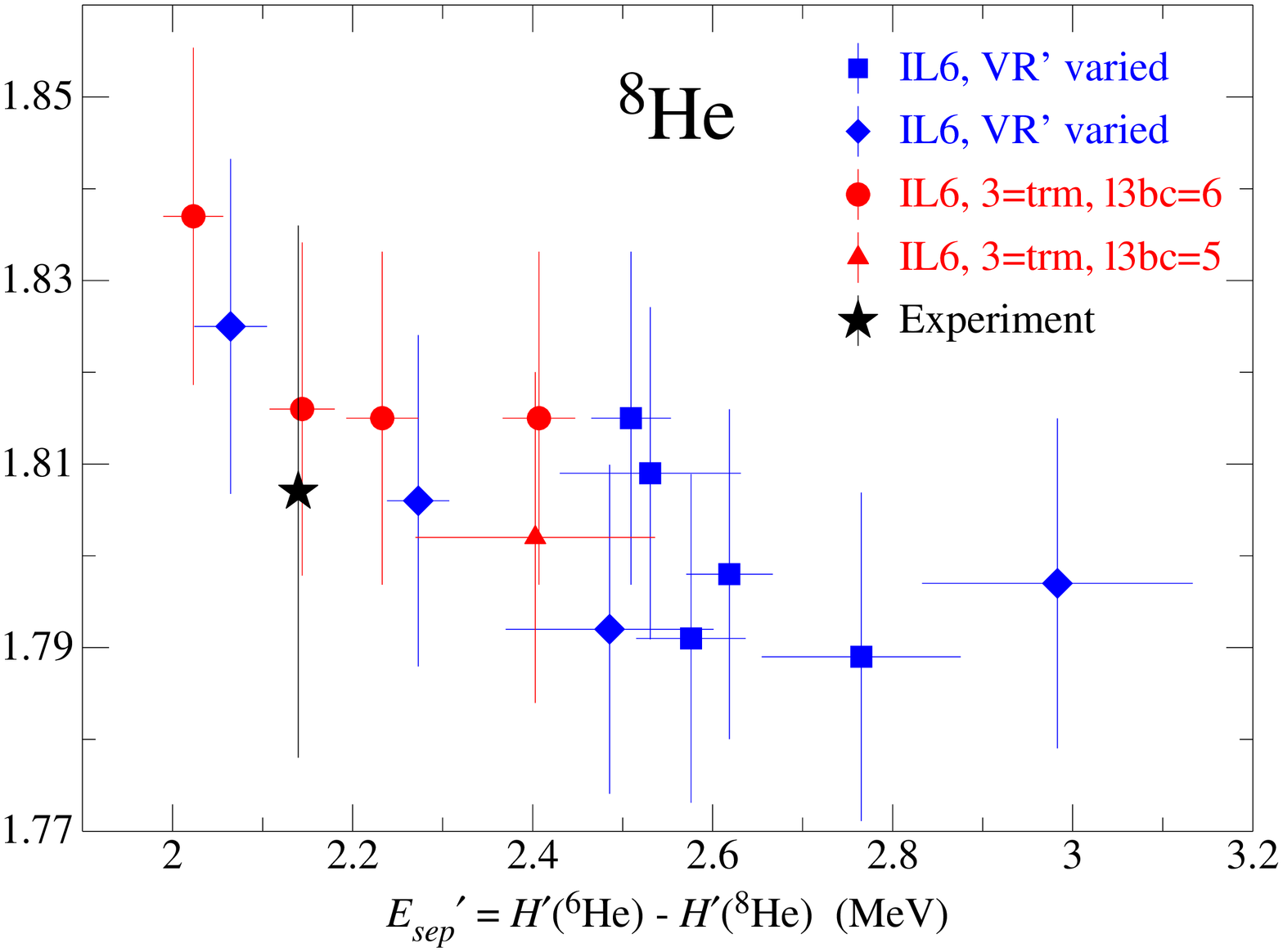}
\end{minipage}
\caption{GFMC calculations of point proton RMS radii of $^{6,8}$He
plotted as a function of the two-neutron separation energy obtained in
the calculation.}
\label{fig:rmsr}
\end{figure}

A few years ago, a group at Argonne measured the RMS charge radius of
the radioactive nucleus $^6$He ($\beta$-decay half-life 0.8 sec.) with
the remarkable accuracy of 0.7\%~\cite{6he-rms} and the corresponding
measurement for $^8$He ($\beta$-decay half-life 0.1 sec.) will be
published soon~\cite{8he-rms}.  This has led us to attempt equally precise GFMC
calculations of the corresponding point proton RMS radii.  Such
calculations are very difficult because of the small separation energies
of the two valence neutrons in these isotopes ($E_{\mbox{sep}} = $
0.97~MeV for $^6$He and 2.14~MeV for $^8$He).  Changes in the starting
$\Psi_T$ and other aspects of the GFMC calculations can result in
changes of 200~keV (400~keV for $^8$He) in the computed energy (and
hence $E_{\mbox{sep}}$).  The RMS radius depends strongly on
$E_{\mbox{sep}}$; as $E_{\mbox{sep}}$ goes to zero, the radius goes to
infinity.  Thus we cannot give a precise value for the computed RMS
radius for a specific Hamiltonian.  Instead we find that the computed
values for the same Hamiltonian with different GFMC calculations, or
even for different Hamiltonians, all lie in a band of radius versus
separation energy.  This is shown in Fig.~\ref{fig:rmsr} which
shows results for two Hamiltonians, AV18+IL2 and AV18+IL6, each
with several GFMC calculations (IL6 is a newer, unpublished, version
of IL2).  The stars in
each panel show the experimental point radii at the
experimental separation energies; they are clearly consistent with our
calculations which give 1.92(4)~fm for $^6$He and 1.82(2)~fm, for $^8$He.
These numbers are both significantly bigger than the RMS point
radius of $^4$He which is 1.46~fm.

\begin{figure}[bt] 
\begin{minipage}[t]{2.40in}
\includegraphics[height=2.4in,angle=270]{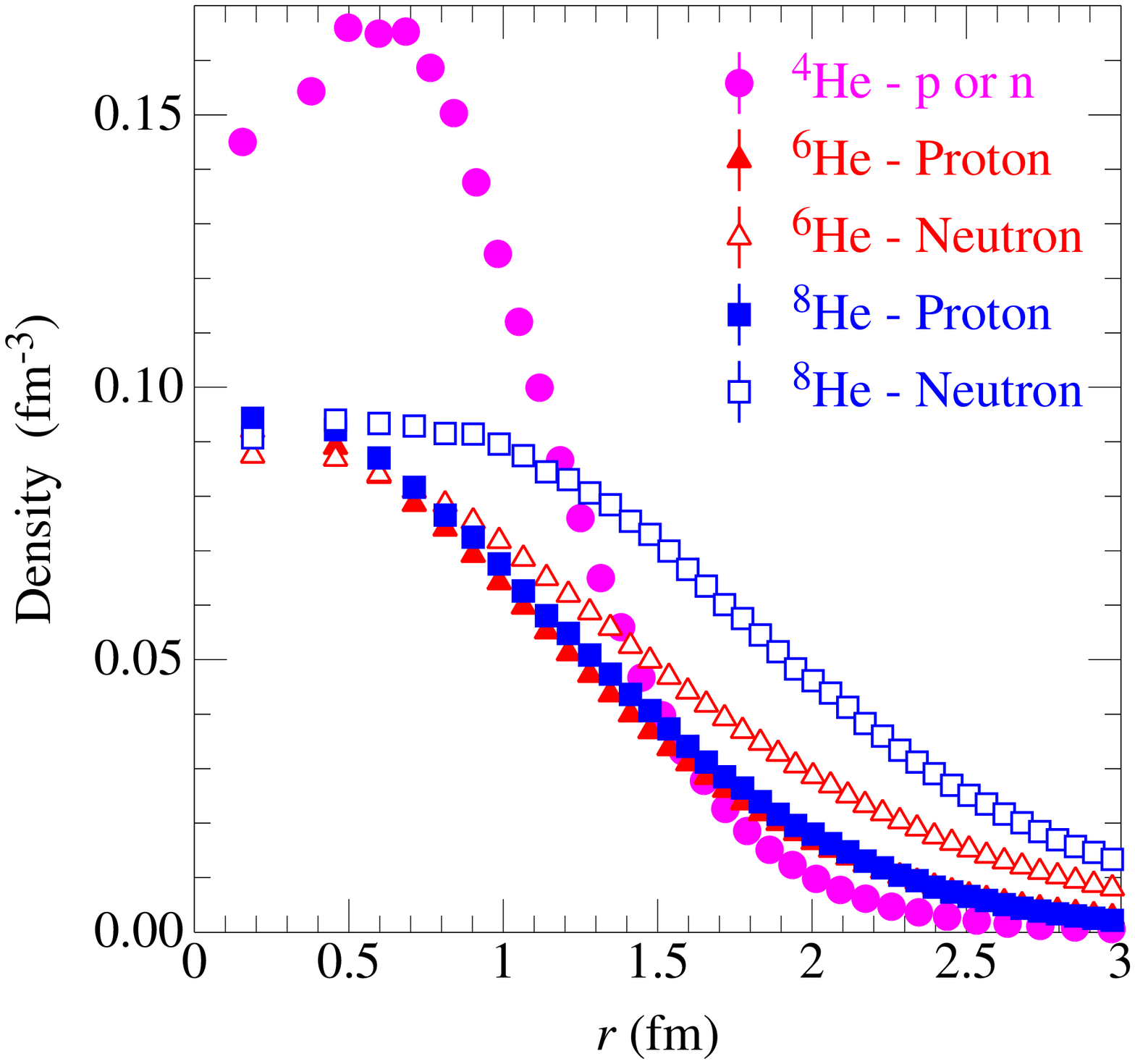}
\end{minipage}
\begin{minipage}[t]{2.50in}
\includegraphics[height=2.9in,angle=270]{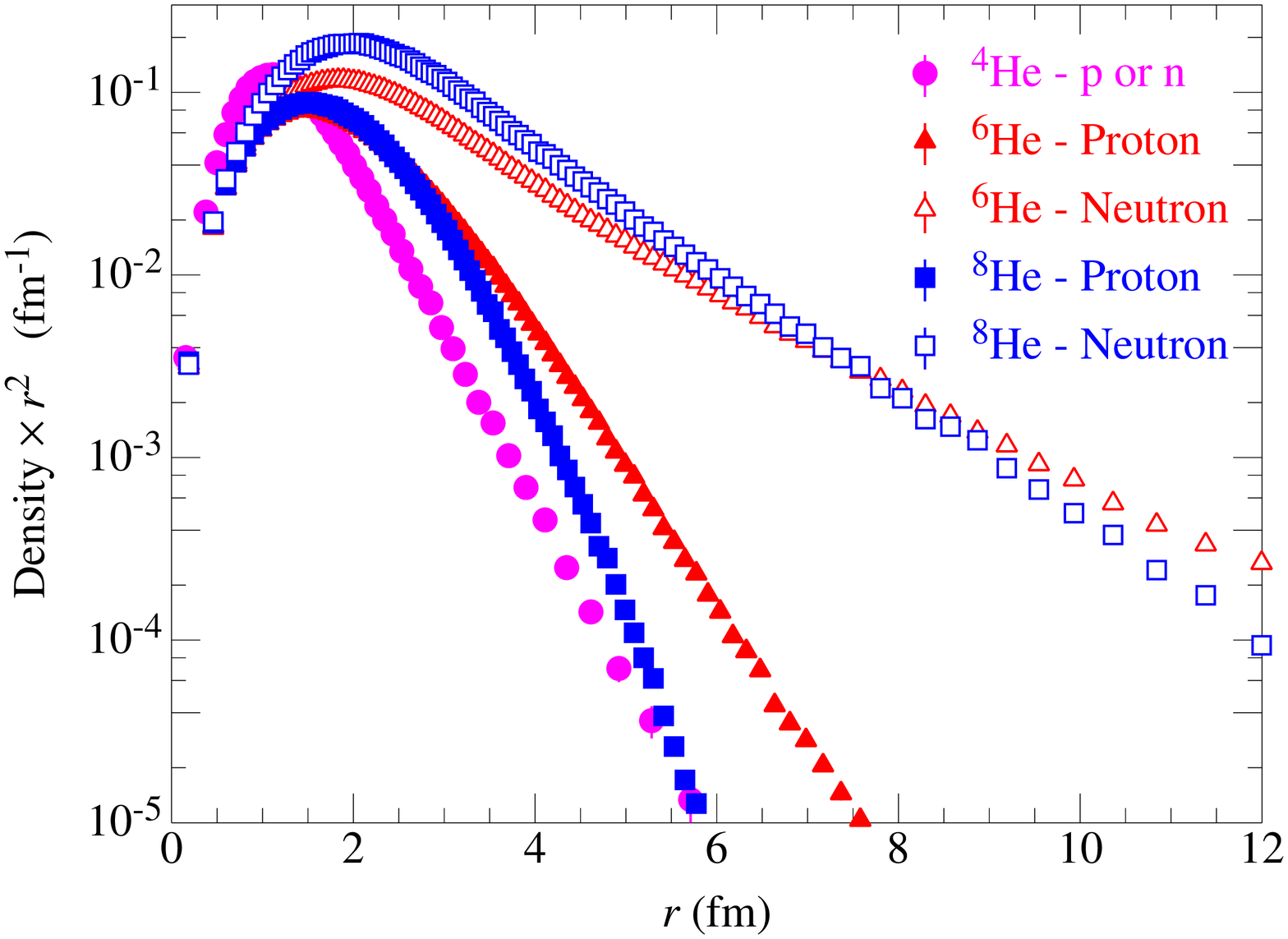}
\end{minipage}
\caption{GFMC calculations, using AV18+IL2, of proton and neutron point
densities for helium isotopes.  The left panel shows density on a linear
scale; the right panel $r^2 \rho$ on a logarithmic scale.}
\label{fig:he-dens}
\end{figure}

Figure \ref{fig:he-dens} shows the point proton and neutron densities of
$^{4,6,8}$He.  The alpha particle is extremely compact; its central
density is twice that of nuclear matter.  In these calculations it has
identical proton and neutron densities which is a very good
approximation.  As is shown below, the valence neutrons in $^{6,8}$He do
not seriously distort the $^4$He core, rather they just drag the $^4$He
center of mass around.  This results in the proton density being spread
out which is why the charge radii of $^{6,8}$He are so much greater than
that of $^4$He even though all three nuclei have just two protons.  The
right panel of the figures clearly shows that $^{6,8}$He have large
neutron halos due to the weak binding of the extra neutrons.  The
neutron halo of $^6$He is more diffuse than that of $^8$He as is
expected from the smaller $E_{\mbox{sep}}$ of $^6$He.


\subsection{Is an alpha particle in a sea of neutrons
still an alpha particle?}

\begin{figure}[bt] 
\includegraphics[height=4.0in,angle=270]{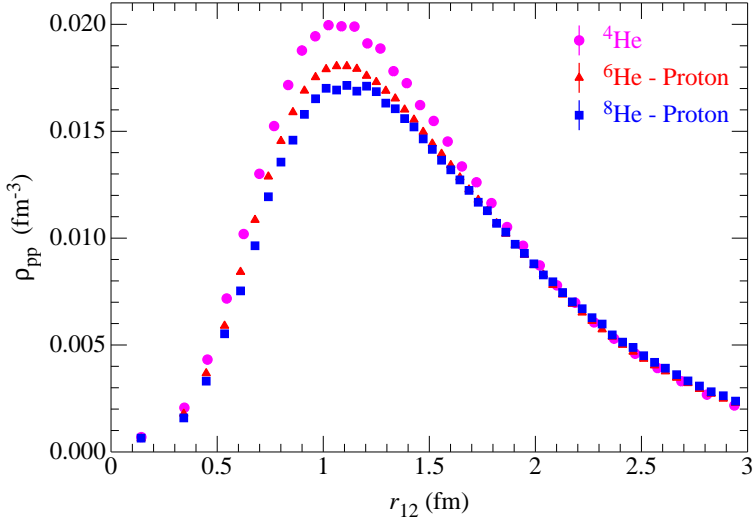}
\caption{GFMC calculations, using AV18+IL2, of two-proton densities of
helium isotopes.}
\label{fig:he-pp}
\end{figure}

The previous subsection showed that the proton density in
$^{6,8}$He is much more spread out than the density of $^4$He,
even though $^{6,8}$He have only extra neutrons added to a $^4$He core.
This might first be thought to indicate that the
core of $^{6,8}$He has been considerably enlarged by the neutrons.
This can be studied by computing  $\rho_{pp}$, the pair density which
is proportional to the probability for finding two protons a given
density apart.
These distribution functions are shown in Fig.~\ref{fig:he-pp}, again
calculated with GFMC for the AV18+IL2 model.
These nuclei each have just one $pp$ pair which presumably is in the
``alpha core'' of $^{6,8}$He.
Unlike the one-body densities, these distributions are not sensitive to
center of mass effects, and thus if the alpha core of $^{6,8}$He is
not distorted by the surrounding neutrons, all three $\rho_{pp}$ distributions
in the figure should be the same.

We see that the $pp$ distribution spreads out slightly with neutron
number in the helium isotopes, with an increase of the pair rms radius of
approximately 4\% in going from $^4$He to  $^6$He, and 8\% to $^8$He.
While this could be interpreted as a swelling of the alpha core, it might
also be due to the charge-exchange ($\tau_i \cdot \tau_j$) correlations
which can transfer charge from the core to the valence nucleons.
Since these correlations are rather long-ranged, they can have
a significant effect on the $pp$ distribution.
VMC calculations of $^4$He with wave functions modified to give
$\rho_{pp}$ distributions close to those of $^{6,8}$He suggest
that the alpha cores of $^{6,8}$He are excited by $\sim80$ and $\sim350$ keV,
respectively, which corresponds to only a 0.4$-$2\% admixture
of the first 0$^+$ excited state of $^4$He at 20~MeV.  Thus
almost all of the increased RMS radius of the proton density is
due to the $\alpha$ core of $^{6,8}$He being pushed around by the neutrons
and not distortions of the core.


\subsection{Two-nucleon knockout -- $(e,e^{\prime}pN)$}

\begin{figure}[bt] 
\includegraphics[height=3.00in]{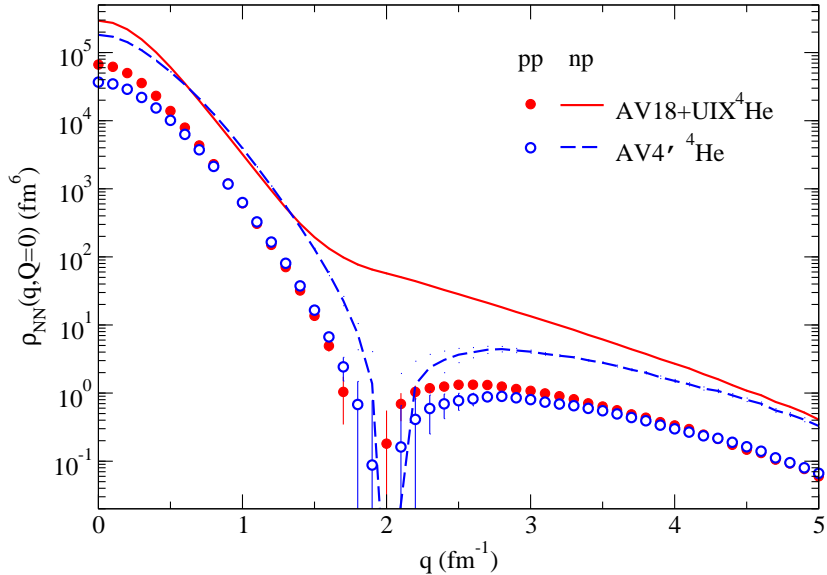}
\caption{Two-nucleon momentum distributions in $^4$He,
computed by VMC for the AV18+UIX Hamiltonian.  The symbols
show $pp$ distributions; the curves are for $np$.}
\label{fig:rho-Pp-4He}
\end{figure}

A recent JLAB experiment for $^{12}$C($e,e^{\prime}pN)$ measured back to
back $pp$ and $np$ pairs; that is pairs with total C.M. momentum $Q=0$,
as a function of their relative momentum, $q$~\cite{Ppexpt}.  They found
that the cross section for $np$ pairs with $q$ in the range 2--3
fm$^{-1}$ is 10--20 times larger than that for $pp$ pairs in the same
range.  To study this we made VMC calculations of the corresponding pair
momentum distributions ($\rho_{\mbox{NN}}$) in several nuclei from $^3$H
to $^8$Be~\cite{SWPC07}.  The calculations for $^4$He are shown in
Fig.~\ref{fig:rho-Pp-4He}.  Results for the AV18+UIX Hamiltonian (which
for $A$=3,4 is a good approximation to AV18+IL2) are shown as the solid
line ($np$ pairs) and solid circles ($pp$ pairs).  Around
$q$=2~fm$^{-1}$ there is a deep minimum in the $pp$ density which
results in large values for $\rho_{np}/\rho_{pp}$.  The dashed curve and
open circles show the corresponding quantities computed for the
AV4$^\prime$ \NN potential (Sec.~\ref{sec:avx}) with no \NNN potential;
in this case both densities have a deep dip and there is no enhancement
of the ratio.

The AV4$^\prime$ potential has no tensor force and thus $np$ pairs are
just $S$-wave while, for the full Hamiltonian, isospin-0 $np$ pairs,
like the deuteron, have a $D$-wave admixture.  The $S$-wave deuteron
momentum distribution has a zero at 2~fm$^{-1}$ which is filled in by
the $D$-wave contribution, the same as is seen here for $np$ pairs in
$^4$He.  The tensor force is much smaller in the pure $T$=1 $pp$ pairs,
so the deep $S$-wave minimum is not filled in for those pairs, even with
the AV18+UIX Hamiltonian.  Calculations for $^3$He, $^6$Li, and $^8$Be
all show this effect although the deep minimum is somewhat filled in for
$^8$Be.  Thus the JLAB experiment shows the importance of tensor
correlations up to $>$ 3 fm$^{-1}$.

\section{Conclusions}
\label{sec:conclusions}

Quantum Monte Carlo methods are powerful tools for studying light nuclei
with realistic nuclear interactions.  Calculations of $A$ = 6 -- 12
nuclear energies with accuracies of $1-2\%$ are possible and the
AV18+IL2 reproduces binding energies with an average error of order 0.7
MeV for $A = 3-12$.  The \NNN potential is required for overall
$P$-shell energies and for spin-orbit splittings and several level orderings.

The QMC methods allow matrix elements of many operators of interest to
be computed.  This contribution presents rms radii, one- and
two-body densities and two-body momentum distributions.  These are
generally in good agreement with experiment.  Recently GFMC values of
$A$=6,7 electromagnetic and weak transitions have also been computed;
these improve on older VMC calculations and also generally agree with
experiment~\cite{PPW08}.  Another topic not covered here is overlap
functions and the related spectroscopic factors; these are used as input
to calculations (such as distorted-wave Born approximation) of nuclear
reactions; recent results are presented in Refs.~\cite{W+PW05,W+PW05b}.
A just-finished interesting study used VMC calculations to investigate
the effects on nuclear binding energies of changes in the Hamiltonian
induced by changes of the fundamental constants~\cite{FW07}.

GFMC calculations of are very computer intensive and at present $^{12}$C
can just barely be done.  However a new generation of extremely parallel
computers is becoming available and we are working with computer
scientists to enable the GFMC program to make use of these machines.
This should lead to the possibility of detailed studies of $^{12}$C
including second 0$^+$ (Hoyle) state which is the doorway for
triple-alpha burning.  This state has resisted precise calculation by
shell-model based methods; we hope that our more flexible variational
wave functions, combined with GFMC propagation, will overcome these difficulties.

But perhaps the most important advance in nuclear GFMC is the
computation of scattering states.  In these calculations the correct
scattering-wave boundary condition is achieved.  The resulting wave
functions will be used to compute reactions of astrophysical interest
such as $^3$He+$\alpha$ $\rightarrow ^7$Be, p+$^7$Be $\rightarrow ^8$B,
and n+($\alpha$+$\alpha$) $\rightarrow ^9$Be.  Indeed all big-bang
nucleosynthesis, solar neutrino, and some $r$-process seeding
reactions are accessible.


\acknowledgments

As can be seen from the author lists of the citations, the work reported
in this contribution is the result of long-term collaborations with
Joesph Carlson (who invented nuclear GFMC), Kenneth M. Nollett (who is
doing the GFMC scattering), Vijay R. Pandharipande (who for many years
guided and inspired our group), Rocco Schiavilla (who has provided the
expertise on the electroweak currents) and Robert B. Wiringa (who has
developed the VMC wave functions).  I thank R.~B.~Wiringa for also
making a critical reading of this MS.  The work would not have been
possible without extensive computer resources provided over the years by
Argonne's Mathematics and Computer Science Division (most recently on
the IBM Blue Gene), Argonne's Laboratory Computing Resource Center, and
the U.S. Department of Energy's National Energy Research Scientific
Computing Center.  This work is supported by the U.S. Department of
Energy, Office of Nuclear Physics, under contract DE-AC02-06CH11357.


\bibliographystyle{varenna}
\bibliography{bibliography}


\end{document}